\documentclass[aps,pra,twocolumn,floatfix,tightenlines,showpacs,notitlepage]{revtex4-2}
\usepackage{amsfonts} 
\usepackage{amsmath} 
\usepackage{amssymb}
\usepackage[normalem]{ulem}
\usepackage{blindtext}
\usepackage{lineno}
\usepackage{cancel}
\usepackage{bm}
\usepackage[usenames, dvipsnames]{color} 
\usepackage{dcolumn}
\usepackage{epic} 
\usepackage{epsfig}
\usepackage{eucal} 
\usepackage{esdiff}
\usepackage{color}
\usepackage{graphicx}
\usepackage[dvipsnames]{xcolor}
\definecolor{darkgreen}{HTML}{005A00} 

\usepackage[breaklinks,colorlinks = true,linkcolor = blue,urlcolor=blue,citecolor=blue]{hyperref}

\begin{document}
\title{Extreme dynamics and relaxation of quantum gases: A hydrodynamic approach}
\author{Ritwik Mukherjee} 
\email{ritwik.mukherjee@icts.res.in}
\author{Abhishek Dhar}
\email{abhishek.dhar@icts.res.in}
\author{Manas Kulkarni}
\email{manas.kulkarni@icts.res.in}
\author{Samriddhi Sankar Ray} 
\email{samriddhisankarray@gmail.com}
\affiliation{International Centre for Theoretical Sciences, Tata Institute of Fundamental Research, Bengaluru 560089, India}
\begin{abstract} 

The evolution of quantum gases, released from traps, are studied through hydrodynamics, both  analytically and numerically, in one and two dimensions. In particular, we demonstrate the existence of long time self-similar solutions of the Euler equations, for the density and velocity fields, and derive the scaling exponents as well as the scaling 
functions. We find that the expanding gas develops a shock front and the size of the cloud grows in time as a powerlaw. 
We relate the associated exponent to that appearing in the 
corresponding equation of state of the quantum gas. Furthermore, we study the relaxation dynamics of a trapped quantum gas and show that the resulting steady state is in excellent agreement with that derived analytically. Our hydrodynamic approach is versatile and can be used to unravel several other far-from-equilibrium collective phenomenon of extreme nature, relevant to the growing experimental interests in quantum gases. 

\end{abstract}
\date{\today}
\maketitle

\section{Introduction}
The non-equilibrium dynamics of cold atomic gases has been a subject of great interest recently both experimentally and theoretically~\cite{Bloch2012,Adams_2012,Bulgac2012,annurev2015,proukakis2013quantum,langen2015non,mr_ed_review,DEFENU20241,langen-natphys,Chomaz_2023}.  Fascinating experiments have been performed~\cite{o2002,menotti2002,kinoshita2004,JTKA2011,ernst1998free} to observe the free expansion and relaxation of trapped quantum gases and there has been significant progress in their theoretical understanding through the framework of hydrodynamics~\cite{Salasnich_2011,kulkarni2012} and generalized hydrodynamics~\cite{ruggiero2020}. While many of the observed experimental features can be explained by Euler hydrodynamics, the role of viscosity and its quantum aspects has been explored in experiments~\cite{cao2011universal,Wang-prl-2022,xiang-2024} and some features are now understood from theory~\cite{Bruun2005,gabriel-prl-2012,trivedi-2017}. However, the exact form of the viscous dissipation in the hydrodynamic equation is far from known.

\begin{figure*}
\includegraphics[width=0.32\linewidth]{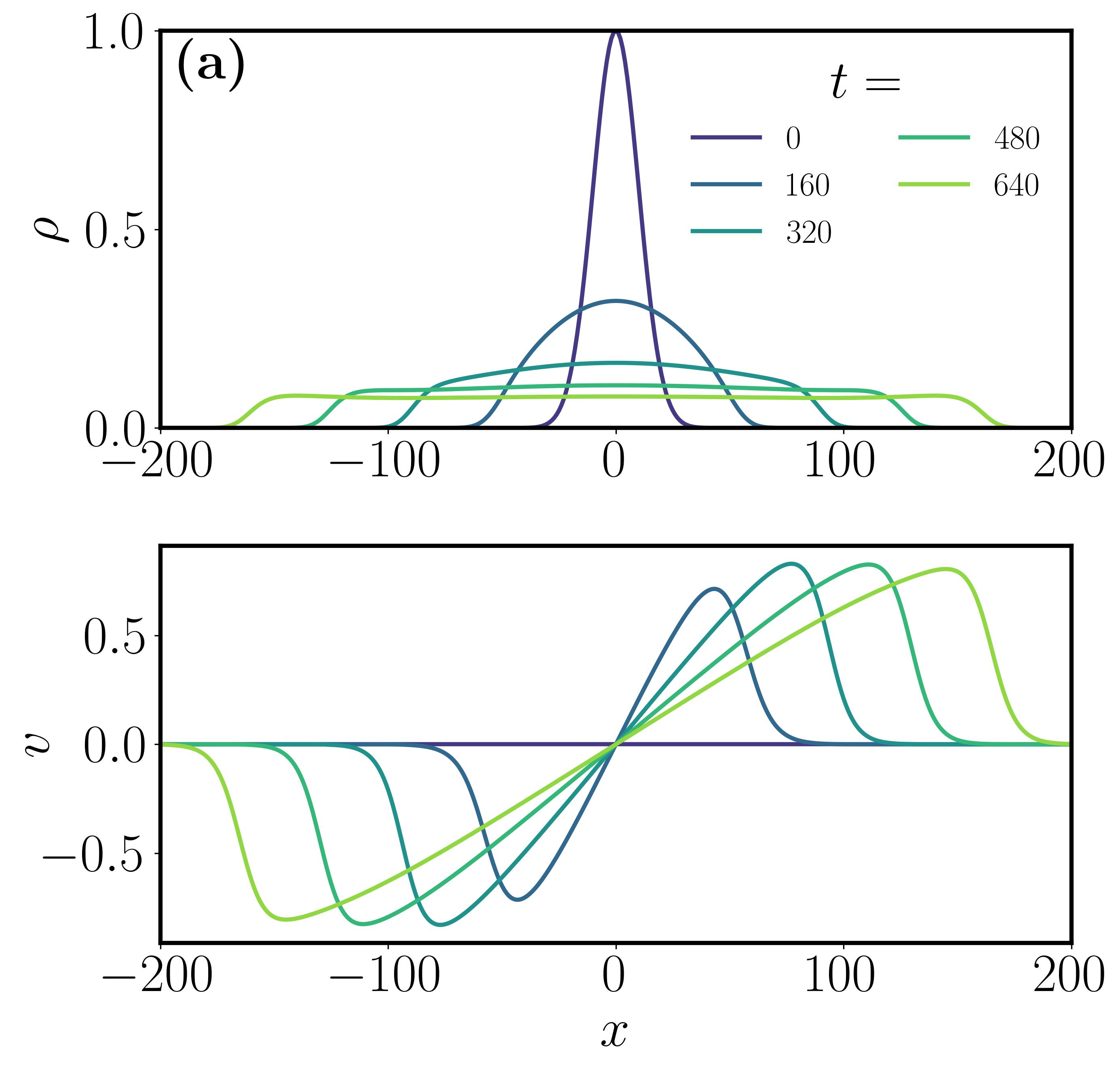}
\includegraphics[width=0.32\linewidth]{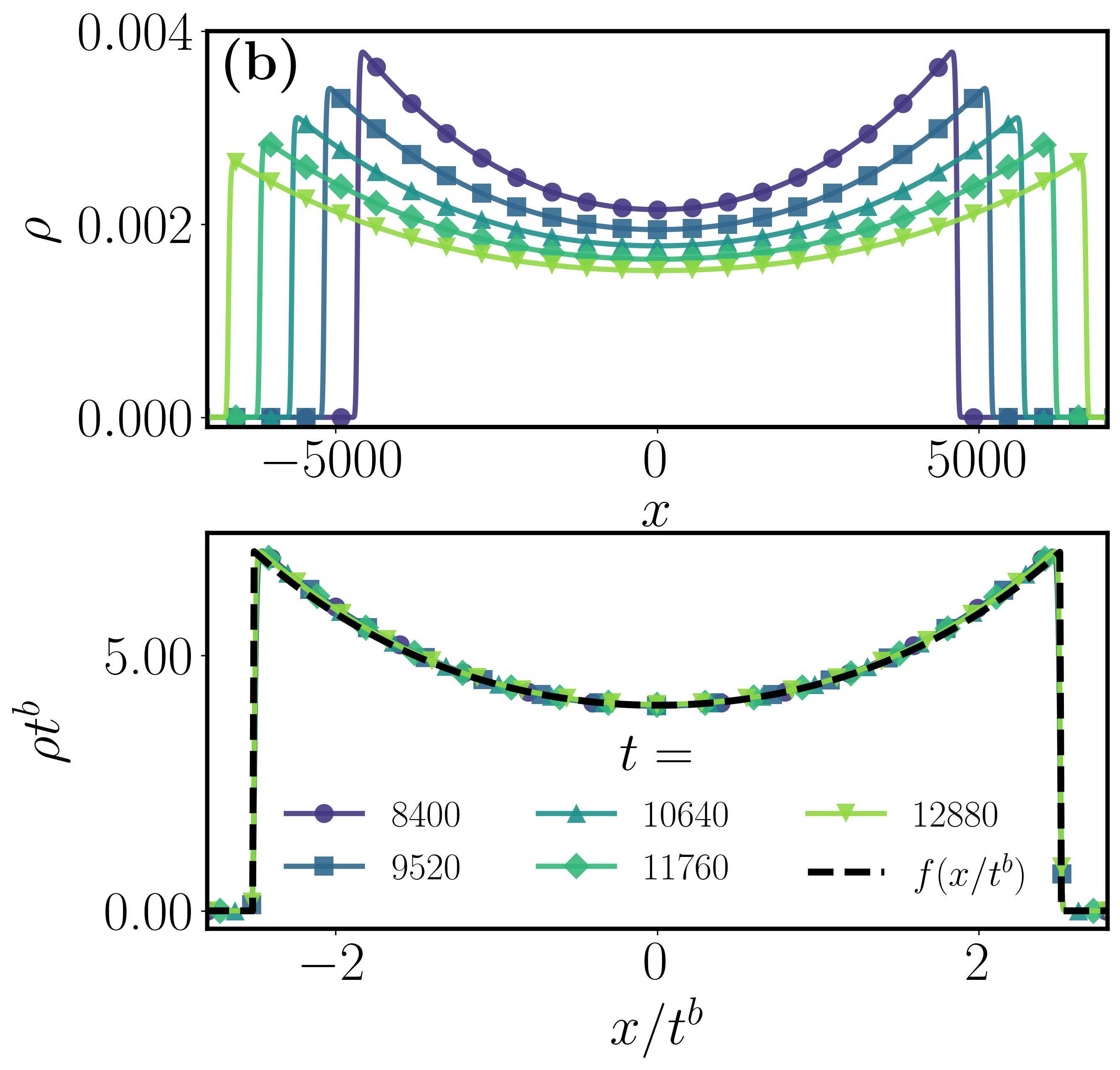}
\includegraphics[width=0.32\linewidth]{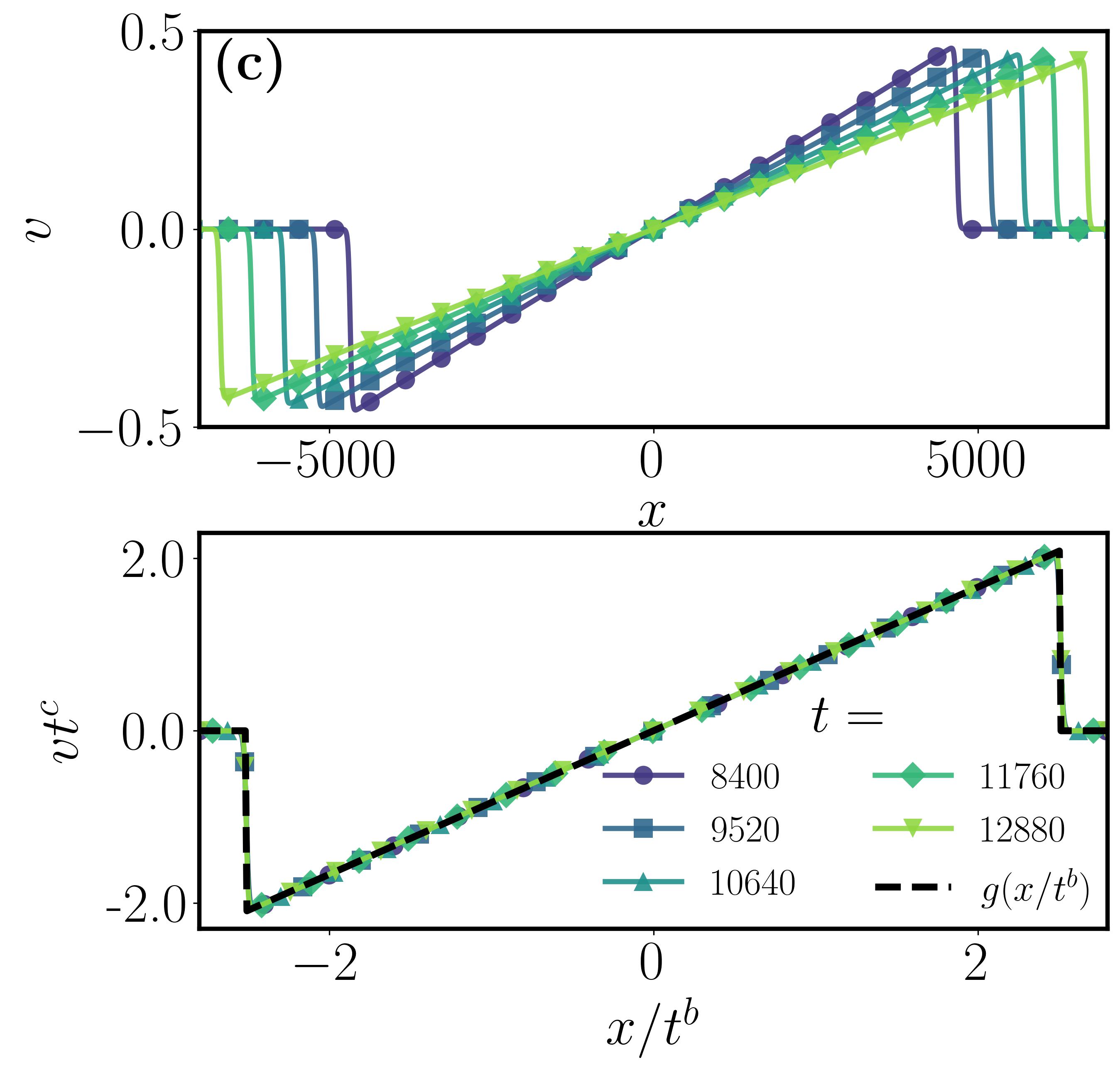}
	\caption{
(a) Time evolution of the density and velocity profiles of a freely expanding 1D quantum gas with $\delta = 2/5$, shown at early times before the emergence of scaling behavior. 
(b) The upper panel shows the late-time evolution of the density profile. The lower panel displays rescaled density plots, $t^b \rho$ versus $\xi = x/t^b$, exhibiting excellent data collapse and precise agreement with the exact scaling function (black dashed line) given by Eq.~\eqref{eq:main:scaling1}. 
(c) The upper panel shows the velocity profile at late times, while the lower sub-panel presents the rescaled velocity profiles, which shows fantastic data collapse and perfect agreement with the exact scaling function described by Eq.~\eqref{eq:main:scaling2}.
} 
\label{fig:1Dprofile}
\end{figure*}

In the context of dynamics in quantum gases, there has been immense activity in bosons~\cite{Rigol_RevModPhys}, fermions~\cite{inguscio2008ultra} and various mixtures~\cite{bose_mixture_science,Semeghini2018,Sowiński_2019}. Such dynamics shines light on the consequences of the underlying interactions that lead to exotic collective phenomenon. There has been studies in understanding dynamics near an equilibrium background (such as sounds waves~\cite{PRL_sound_2006,PRL2010_sound,PRL_sound,Kuhn2020,Drummond18,Li_Science_22,martin-science}) and dynamics very far from equilibrium (for e.g., resulting in shock waves~\cite{JTKA2011,kulkarni2012,LS2013, BulgacPRL} or energy dynamics~\cite{yan-arxiv}).
However, in these systems dynamics of extreme nature is far less understood. One such
specific problem is to study an evolution that mimics a
blast that can arise from a sudden release of a substantial amount of particles or energy from a localized
region. Some natural challenges while studying these problems include: (i) establishing  rigorously the  collective hydrodynamic description (ii) unavailability of analytical solutions, especially very far from equilibrium even if a phenomenological hydrodynamic theory is available. 

At the level of Euler hydrodynamics, one of the most interesting results are the so-called self-similar scaling solutions~\cite{sedov2018}, first discussed by Taylor, von Neumann and Sedov~\cite{sedov1946,Taylor50a,Taylor50b,vN} in the context of atomic explosions.  The main question there is to understand the time evolution of the hydrodynamic fields, namely density, momentum and energy, for an initial state where a huge amount of localized energy is released in an otherwise cold ambient gas --- this is referred to as the blast problem. Following the classic work of TvNS, there has been much recent interest in the blast problem, in particular looking at direct comparisions of the Euler predictions with results from microscopic molecular dynamics simulations and understanding the effects of dissipation~\cite{Antal2008,Barbier2015,Barbier2015a,Rajesh2021a,Rajesh2021b,Chakraborti_PRL_2021,ganapa2020taylor,singh2023}. So far, all work has focused on classical systems and the main goal of the present paper is to examine self similar solutions for blasts in  quantum gases. Furthermore we consider quantum gases confined in traps and study the approach to equilibrium using the framework of hydrodynamics.  

While the physics of blasts is rooted in the fascinating,
collective dynamics of the constituent bosons or fermions, it is possible to
capture the essential, experimentally measurable dynamics through a
coarse-grained hydrodynamic description.  Here we consider the zero temperature case where it is sufficient  to consider the equations for the density field, $\rho({\bf x},t)$, and velocity field, ${\bf v}({\bf x},t)$. The hydrodynamic Euler equations, in the presence of an external trapping potential, $U_{\rm trap}({\bf x})$,  are given by
\begin{subequations}
\label{eq:eulergen}
\begin{align}
\label{eq:cont}	
	&\frac{\partial \rho}{\partial t} + \nabla\cdot(\rho {\bf v}) = 0, \\
    &\frac{\partial {\bf v}}{\partial t} + {\bf v}\cdot\nabla {\bf v} + \nabla w = -\nabla U_{\rm trap},
\label{eq:euler}	
\end{align}
\end{subequations}
 where the specific enthalpy $w$ is related to the pressure, $P$, via the relation $\partial_\rho P=\rho \,\partial_\rho w$. A knowledge of the zero-temperature equation of state of the quantum gas thus enables us to write the hydrodynamics equations. 
Interestingly, a large number of quantum gases, both
bosonic as well as fermionic, have a simple power-law form for the specific enthalpy 
\begin{equation}
w(\rho) = D\,\rho^\delta.
\label{eq:wrho}
\end{equation}
As examples in one-dimensional ($d=1$) systems, we have: $\delta = 1$ and $\delta=2$ for the Lieb-Liniger gas of bosons at weak and strong coupling respectively; 
$\delta = {2}/{5}$ is the limiting case of a quasi-1D (cigar-shaped) unitary Fermi gas~\cite{JTKA2011,kulkarni2012}; $\delta=2/3$
for unitary Fermi gas in plane-wave limit. 
In dimensions $d>1$ examples include: $\delta=2/d$ for weakly interacting fermions at zero temperature. On the other hand, the unitary Fermi gas in $d$ dimensions  also shares the same dependence on $\delta$. However, there is an overall renormalization in comparison to weakly interacting case~\cite{JTKA2011,LS2013}. Remarkably, the unitary fermi gas, being a strongly correlated quantum fluid, shares deep resemblance to very different systems such as the quark–gluon plasma~\cite{SHURYAK200948, Schäfer_2009,Adams_2012} thereby enabling a hydrodynamic description.

It is worth noting that, beyond cold gases, the well-known Riesz family of gases~\cite{ML2022,SA2019,kumar-frg,santra-prl,kiran2021} provides an important and distinctly complementary example described by Eqs.~(\ref{eq:eulergen},\ref{eq:wrho}). This family comprises of particles with all-to-all interactions of power-law type characterized by an exponent, say $s$. The collective theory of the Riesz family of gases in a broad range of parameter space falls under the general structure of Eqs.~\eqref{eq:cont} and \eqref{eq:euler}. In particular, a class of these models are of finite-ranged type~\cite{kumar-frg}. It was shown in Ref~\onlinecite{kumar-frg} that a finite-ranged version of the Riesz gas in one dimension characterised by exponent $s$, would correspond to $\delta=s$ [Eq.~\eqref{eq:wrho}] for any $s>0$. This immediately offers us a plethora of possibility of realization of the same collective dynamics for a wide window of $\delta$.

The precise set-up where we study blasts in quantum gases is motivated by experiments in cold atoms using, e.g, time-of-flight (TOF) techniques, a standard tool in atomic physics for probing momentum and coherence in quantum gases. In TOF experiments, atoms are cooled in a (typically harmonic) trap, abruptly released, and  allowed to expand for a set time, and then imaged via absorption imaging techniques~\cite{abs_img_1, abs_img_2,abs_img_3}. In the present work we study, through Euler hydrodynamics~(Eqs.~\ref{eq:eulergen}), the dynamical evolution of such gases after their release from the trap. We show that, at long times, they evolve into self-similar forms and provide a complete analytic and numerical understanding. Secondly, we also consider the problem of a quantum gas in a  trap
($U_{\rm trap} \neq 0$) and study the relaxation dynamics to the  steady state distribution. In this case, we add the Navier-Stokes corrections to Eqs.~\eqref{eq:eulergen}.  For reasons explained later, we restrict ourselves to $0<\delta<2$ for the one dimensional case and $0<\delta<1$ for the two dimensional case.

This paper is organized as follows. 
In Sec.~\ref{sec:free_1D}, we discuss the one-dimensional gas without any external trap. Both evolution in zero (Sec.~\ref{sec:vac}) and finite background (Sec.~\ref{sec:bg}) are discussed.  We then discuss the relaxation dynamics of one-dimensional gas in a trap in Sec.~\ref{sec:trap}. The two dimensional case is discussed in Sec.~\ref{sec:2d}. We end with conclusions along with an outlook in Sec.~\ref{sec:conc}. Details of direct numerical simulations are provided in Appendix.~\ref{sec:dns}.

\section{Free evolution in one-dimensional case} 
\label{sec:free_1D}
We begin with the one-dimensional (1D) problem with the trap switched off and consider the Euler equations for the density and velocity fields:
\begin{subequations} 
		\label{eq:1Deuler}
	\begin{align}
\frac{\partial \rho}{\partial t} + \frac{\partial \rho v}{\partial x} &= 0,
		\label{eq:1Deulerb}
		\\
\frac{\partial v}{\partial t} + \frac{\partial }{\partial x}\left (\frac{v^{2}}{2}+D \rho^\delta \right) &= 0,
		\label{eq:1Deulera}		
	\end{align}
\end{subequations} 
We consider initial condition where we start with a density profile on top of a uniform density background $\rho_B$ and zero velocity field. The cases with background density either zero and non-zero  will be considered separately in the following sections. Let us assume that the density profile has a spatial spread over a characteristic length $\sigma$ and a typical size $\rho_0$.  We can then  transform to dimensionless variables 
\begin{align}
\label{eq:DL}
\rho/\rho_0 \to \rho,\, x/\sigma \to x, ~ \frac{v}{ (D\rho_0^\delta)^{1/2}} \to v,~ \frac{t(D\rho_0^\delta)^{1/2} }{\sigma} \to t.
\end{align}
This corresponds to setting $D=1$ in Eq.~\eqref{eq:1Deulera} and we will henceforth assume this.

We will consider below the two case where an excess of particles is introduced in a localized region,  either in an empty region (Sec.~\ref{sec:vac}) or one with a finite density of particles (Sec.~\ref{sec:bg}).

\subsection{Evolution in a vacuum}
\label{sec:vac}
In this section, the initial condition  is a spatially localized mass (with centre of mass at the origin) of a gas with $\rho_B=0$ and net zero momentum. It is expected that the Euler equations should, at long times, develop a shock and  the hydrodynamic fields  have  scaling forms. 
At long times  the fields $\rho,v$ will therefore take the forms 
\begin{align}
\rho (x,t)= t^{-b} \, f(\xi),\,
	v (x,t) = t^{-c} \, g(\xi)\,, 
    {\rm where}~~	\xi =\frac{x}{t^b}\, , 
	\label{main:eq:fg}
\end{align}
$f$ and $g$ are scaling functions and $b$ and $c$ are scaling exponents. Plugging the forms in Eq.~\eqref{main:eq:fg} into Eq.~\eqref{eq:1Deulera} and requiring that the resulting  equations in $f(\xi)$ and $g(\xi)$ have no explicit time-dependence fixes the scaling exponents $b$ and $c$ as 
\begin{equation}
	b = \frac{2}{\delta +2 }, \quad \quad c = \frac{\delta}{b+\delta}.
	\label{eq:app:bc}
\end{equation}
We then get the following ordinary differential equations (ODEs) for the scaling functions $f$ and $g$:
\begin{subequations} 
	\begin{align}
		\frac{d (fg)}{d\xi} - b\xi\frac{df}{d\xi} - bf &= 0; 
		\label{eq:g}
		\\
		\frac{df^\delta}{d\xi} + \frac{1}{2}\frac{dg^2}{d\xi} - b\xi\frac{dg}{d\xi} -cg &= 0.
		\label{eq:f}
	\end{align}
	\label{eq:fg}
\end{subequations} 
Let us indicate the location of the shock in scaled variable as $\xi_f$. Therefore, it follows that the position of the shock is given by 
\begin{align}
	R(t)= \xi_f t^b,
\end{align}
 Let us denote the shock velocity 
 \begin{equation}
 U=\dot{R},
 \label{eq:svel}
 \end{equation}and the fields $(\rho_-,v_-)$ and $(\rho_+=0,v_+=0)$ just behind and in front of the shock, respectively. The Rankine--Hugoniot \cite{Chakraborti_PRL_2021,ganapa2020taylor} boundary conditions then give:
\begin{align}
\label{eq:rh}
	v_-=U,~~~\frac{v_-^2/2+ \rho_-^\delta}{v_-}=U, 
\end{align}
and hence
\begin{align}
	f_- = \left(\frac{\xi_f^2 b^2}{2}\right)^{1/\delta}, ~~~g_- = \xi_f b . 
	\label{eq:main:rh_cond}
\end{align}
With these boundary conditions, we can solve Eq.~\eqref{eq:fg} to get
\begin{subequations}
	\begin{align}
    \label{eq:main:scaling1}
		f(\xi) &=  \left(  \frac{\xi_f^2(2-\delta) + \delta\, \xi^2 }{(\delta+ 2)^2} \right)^{1/\delta},\\
		g(\xi) &= b \,\xi.
        \label{eq:main:scaling2}
	\end{align}
\end{subequations}
The factor $\xi_f$ is fixed by using the mass conservation equation 
\begin{equation}
\int_0^{\xi_f} d\xi f(\xi)=M\, ,
\label{eq:conserve}
\end{equation}
which, for $\delta<2$, gives
\begin{equation}
\label{eq:2f1}
\xi_f= \left(\frac{(2-\delta )^{-1/\delta } (\delta +2)^{2/\delta } M}{\, _2F_1\left(\frac{1}{2},-\frac{1}{\delta };\frac{3}{2};\frac{\delta }{\delta -2}\right)}\right)^{\frac{\delta }{\delta +2}},
\end{equation}
where $_2F_1$ represents the Hypergeometric function. Equation.~\eqref{eq:2f1} shows that we can get an explicit form of shock location $\xi_f$ for $\delta \leq 2$. One can immediately see that the case $\delta=1$ gives
$\xi_f = 3 (M/4)^{1/3}$. 

We now test the scaling solutions in Eqs.~\eqref{eq:main:scaling1} and \eqref{eq:main:scaling2} through direct numerical solution of the hydrodynamics equations. The details of the pseudospectral numerical method employed is provided in Appendix.~\ref{sec:dns}. We chose extremely localised yet generic, initial conditions. For a system of length $L$,
\begin{eqnarray}
\rho(x,0) &=& \frac{1}{2\sigma^2}\exp(-x^2/\sigma^2)\quad \rm{with} \quad \sigma \ll L \\
    v(x,0) &=& 0.
\end{eqnarray} 
In Fig.~\ref{fig:1Dprofile} we show results from direct numerical simulation for the time evolution of density and velocity for $\delta=2/5$. In panel (a), we show relatively early time evolution and the subsequent emergence of the shock. In panels (b) and (c) we show the late time evolution of density and velocity and their anticipated scaling and fit to the analytical predictions for the scaling functions given in Eqs.~\eqref{eq:main:scaling1} and \eqref{eq:main:scaling2}.

\begin{figure*}
\includegraphics[width=0.32\linewidth]{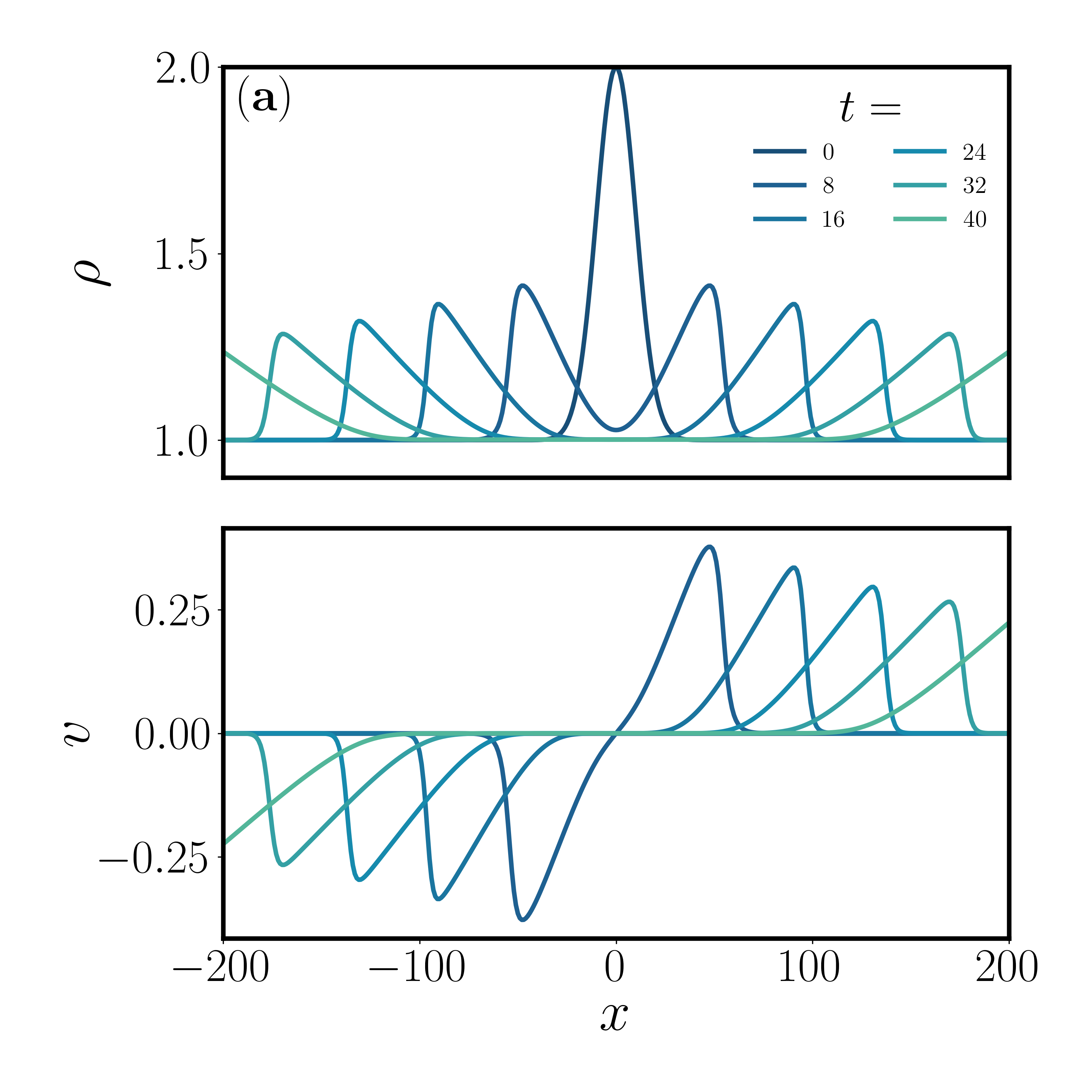}
\includegraphics[width=0.32\linewidth]{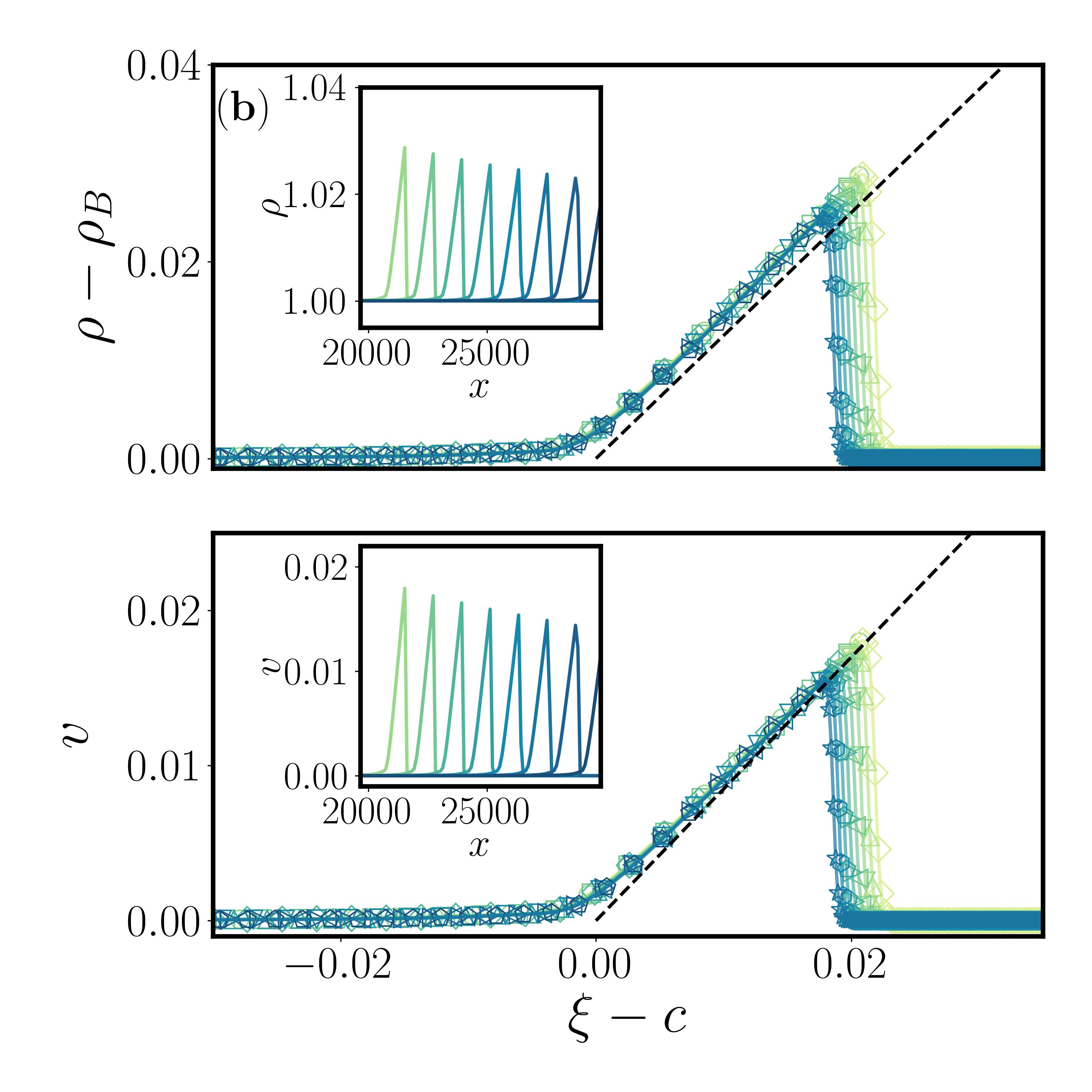}
\includegraphics[width=0.32\linewidth]{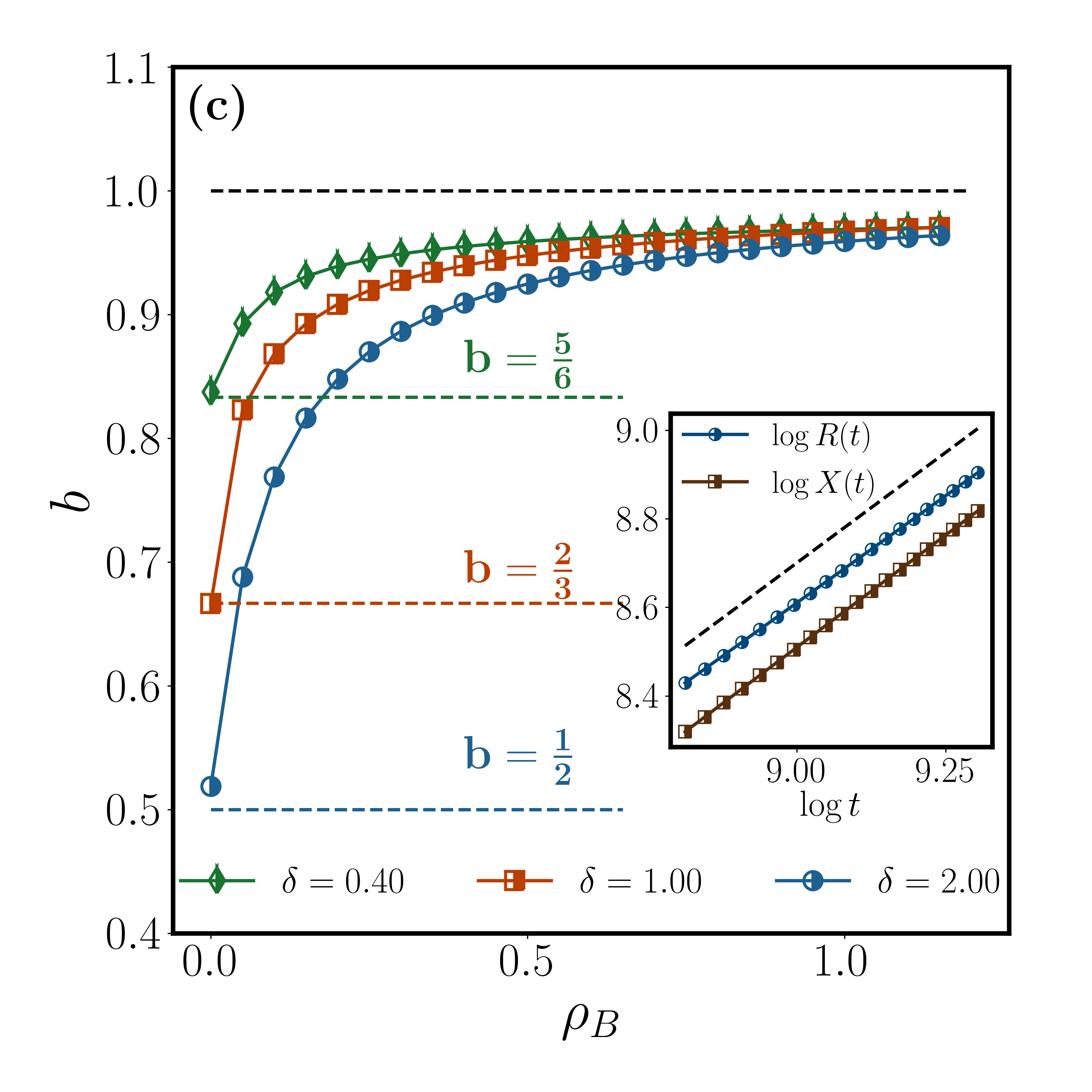}
	\caption{(a) Early time profiles for density and velocity for $\delta = 2/5$ and for background density $\rho_B = 1$ before the development of scaling structure.
    	(b) Long time density and velocity profiles for $\delta = 2/5$ with the insets depicting the density profiles at 
	different times. In the main panel the markers represent $\rho - \rho_B$ at different times 
	with the space rescaled as $\xi = x / t$. The dashed line represents the analytical formula for the density given in Eq.~\eqref{eq:fgBSol}.
    (c) The inset shows the shock location $R(t)$ and ramp location $X(t)$ as a function of time. Dashed line is a guide to the eye sowing the ballistic behaviour. The main panel shows the scaling exponent as a function of background density $\rho_B$ for the same excess density same. As expected $\rho_B \to 0$ the exponents goes to the vacuum exponents.}
\label{fig:bg}
\end{figure*}

\begin{figure}    \includegraphics[width=0.8\linewidth]{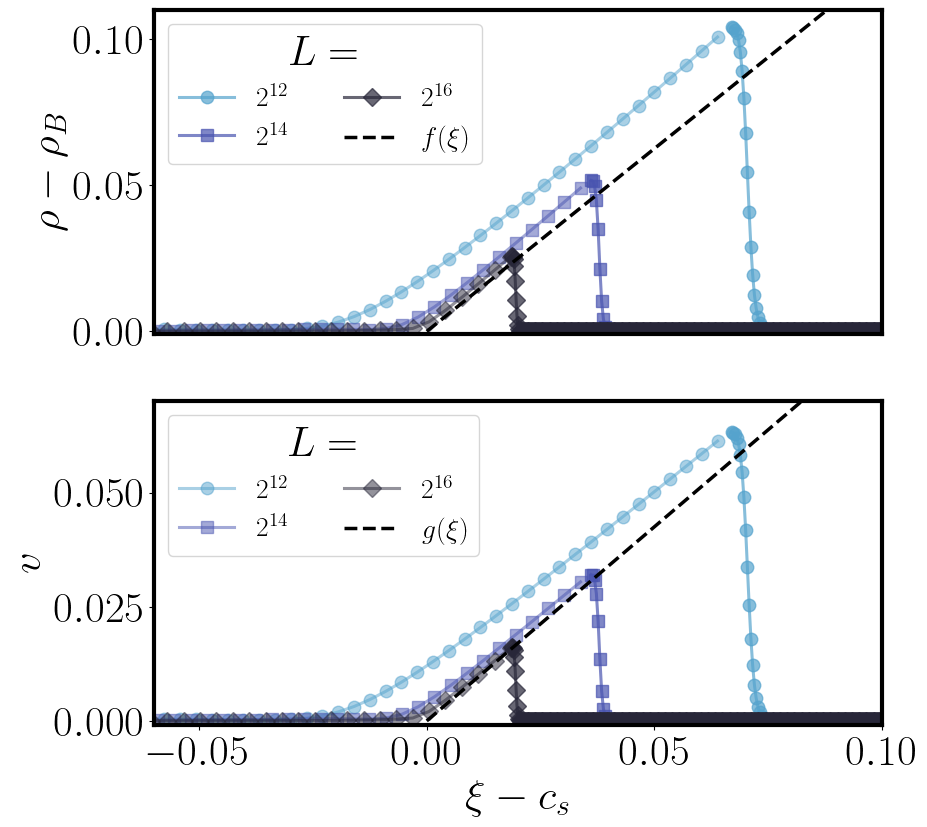}
     \caption{This plot shows the convergence of the rescaled profile to the analytical solution given in Eqs.~\eqref{eq:fBsol} and~\eqref{eq:gBsol} for $\delta = \frac{2}{5}$ and $\rho_B = 1$. As the system size increases, the numerical profile approaches the asymptotic limit and converges to the analytical expression.
}
    \label{fig:convg_rhoB}
\end{figure}

\subsection{Evolution in a finite density background}
\label{sec:bg}

We now consider the case where we add mass to an existing  finite background density $\rho_B$. The localized mass density will then propagate and simultaneously spread. At sufficiently large times, the excess density is a perturbation and one can then do a linearized theory, which would predict the propagation to be ballistic characterized by a speed of sound $v_s$. If the initial excess density is  large then the initial propagation will be sub-ballistic. However, after a considerable time has passed, the  excess density will become small compared to the background thereby validating a linearized theory. The linearized theory can be obtained from Eq.~\eqref{eq:1Deuler} as (setting $D=1$)
\begin{equation} 
\label{eq:1DeulerL}
\frac{\partial \Delta \rho}{\partial t} + \rho_B\frac{\partial  \Delta v}{\partial x} = 0, \quad \frac{\partial \Delta v}{\partial t} + \delta \rho_B^{\delta-1}\frac{\partial }{\partial x}\left ( \Delta \rho \right) = 0,
\end{equation} 
where $\Delta\rho$ and $\Delta v$ denote small perturbations to justify linearization. Equation~\eqref{eq:1DeulerL} immediately gives the sound speed 
\begin{equation}
\label{eq:sound}
c_s = \sqrt{\delta\,\rho_B^\delta}\, .  
\end{equation}
We now discuss the scaling solutions at long times. We expect ballistic behaviour at long times and hence we assume the form 
  \begin{align}
  \label{eq:fb_fg}
      \rho = \rho_{B} + f(x/t),~~v = g(x/t), 
  \end{align}
  where $f(x/t)$ and $g(x/t)$ are scaling functions that we seek to compute analytically. Plugging Eq.~\eqref{eq:fb_fg}  into Eq.~\eqref{eq:1Deuler}	yields
\begin{subequations}
\label{eq:bgmain}
	\begin{align}
    -\xi \frac{df}{d \xi} + \rho_B \frac{dg}{d\xi} + \frac{dfg}{d \xi} &= 0 \\
		-\xi \frac{d g}{d \xi} +g \frac{d g}{d\xi} + \delta (\rho_B
		+f)^{\delta-1} f^{\prime} &= 0\,,
        \label{eq:bg1}
	\end{align}
\end{subequations}
where we recall that $\xi=x/t$. 
Equation~\eqref{eq:bgmain} can be recasted as 
	\begin{subequations}
		\begin{align}
			(g-\xi) \frac{df}{d\xi} + (f+\rho_B) \frac{dg}{d\xi}  &= 0;
		\label{eq:fB} \\
		\delta (f+\rho_B)^{\delta-1} \frac{df}{d\xi} + (g-\xi)\frac{dg}{d\xi}    &= 0.
		\label{eq:gB}
		\end{align}
		\label{eq:fgB}
	\end{subequations}
Equation~\eqref{eq:fgB} is a linear homogeneous equation in $f'$ and $g'$ and therefore a solution exists only if the determinant $(g-\xi)^2-\delta (f+\rho_B)^\delta = 0$. This immediately gives us a relation between $g$ and $f$
\begin{align}
	g= \xi \pm \sqrt{\delta} (f+ \rho_B)^{\delta/2}.
    \label{eq:rel_gf}
\end{align}
Plugging Eq.~\eqref{eq:rel_gf} into  Eq~\eqref{eq:fB} one gets
\begin{align}
	 \pm (1+\delta/2) \sqrt{\delta} (f+ \rho_B)^{\delta/2} f' +(f+\rho_B) &= 0.
\end{align}
This can be integrated to yield (the physical solution)
\begin{subequations}
\label{eq:fgBSol}
	\begin{align}
		\label{eq:fBsol}
		f (\xi) &= \rho_B -\Phi \, \\
		g (\xi) &= 
		\xi  - \sqrt{\delta } \left(2 \rho_B -\Phi \right)^{\delta /2}\,
		\label{eq:gBsol}
	\end{align}
\end{subequations}
where, 
\begin{equation}
\Phi =  \left(\frac{  2 (\delta +1) \text{$\rho_B$}^{\delta /2}-\delta^{1/2}  \xi}{\delta +2}\right)^{2/\delta }.
\end{equation}
Note that, as requried, the following conditions are satisfied
\begin{equation}
f(\xi=c_s) = 0, \quad g(\xi = c_s) = 0\,.
\end{equation}

In Fig.~\ref{fig:bg}(a) we show the density (upper panel) and velocity (lower panel) profiles for different times with the 
	background density $\rho_B = 1$ with $\delta = 2/5$. Clearly, both fields evolve ballistically. After some time, the shock front $R(t)$ moves with the sound speed, i.e.,  $R(t)=c_s t$ where we recall that $c_s$ is given in 
    Eq.~\eqref{eq:sound}. 
    Given the symmetric nature of the evolution, we track the 
	shock moving towards the right. In Fig.~\ref{fig:bg}(b)     
we show the plot in  scaled and shifted variables ($\xi-c_s$) where $\xi = x/t$. In Fig.~\ref{fig:bg}(c) we plot the scaling exponent $b$ as a function of background density $\rho_B$. In the large system size limit, we observe exponent $b=1$
	independent of $\delta$. Clearly, as $\rho_B \to 0$, we recover the freely-evolving gas limit where, as seen earlier 
	$b = \frac{2}{2+\delta}$. In the inset of Fig. 2(c) we plot the  position of shock and the ramp as a function of time. The thick black line, as a guide to the eye, confirms the ballistic 
	motion of $R(t)$. We further analyze the shock front in Fig.~\ref{fig:convg_rhoB}. We demonstrate that as one increases the system size, the numerical profile approaches the asymptotic limit and converges to the analytical expression given in Eq.~\eqref{eq:fgBSol}.  

We next discuss some analytical estimates for the location of the shock and its scaling with time. For this, we revisit the expression of $f(\xi)$ in Eq.~\eqref{eq:fBsol}. Note that for relatively small $\xi-c_s$, we have
\begin{eqnarray}
\label{eq:f_exp}
    f(\xi-c_s) = A (\xi-c_s)+O\big((\xi-c_s)^2\big)\,, 
\end{eqnarray}
where 
\begin{equation}
\label{eq:Aeq}
A = \frac{2}{2+\delta}\frac{\rho_B^{1-\delta/2}}{\sqrt{\delta}}\, .
\end{equation}
Next we require by normalization, 
\begin{equation}
\label{eq:normt}
t\int_{c_s}^{(c_s+x_\text{max}/t)} f(\xi)d\xi = \frac{M}{2}\, ,
\end{equation}
where we recall that $M$ is the excess mass and the factor of $2$ is because we are considering one side of the profile. Simplifying Eq.~\eqref{eq:normt} by using Eq.~\eqref{eq:f_exp} gives
\begin{equation}
    x_\text{max} = \sqrt{\frac{M\,t}{A}}\, . 
\end{equation}
We next evaluate the height of the shock at $\xi = c_s+x_\text{max}/t$. To do so, we plug the value of $x_\text{max}/t$ in Eq.~\eqref{eq:f_exp} to give
\begin{equation}
\label{eq:fmax}
    f(\xi-c_s = x_{\rm max}/t) = \sqrt{\frac{MA}{t}}.
\end{equation}
We have numerically that verified that  $f(\xi-c_s = x_{\rm max}/t) \sim 1/\sqrt{t}$. In fact, our numerical results are also consistent with the precise coefficient $\sqrt{M\,A}$ given in Eq.~\eqref{eq:fmax}.

\begin{figure*}
\includegraphics[width=0.32\linewidth]{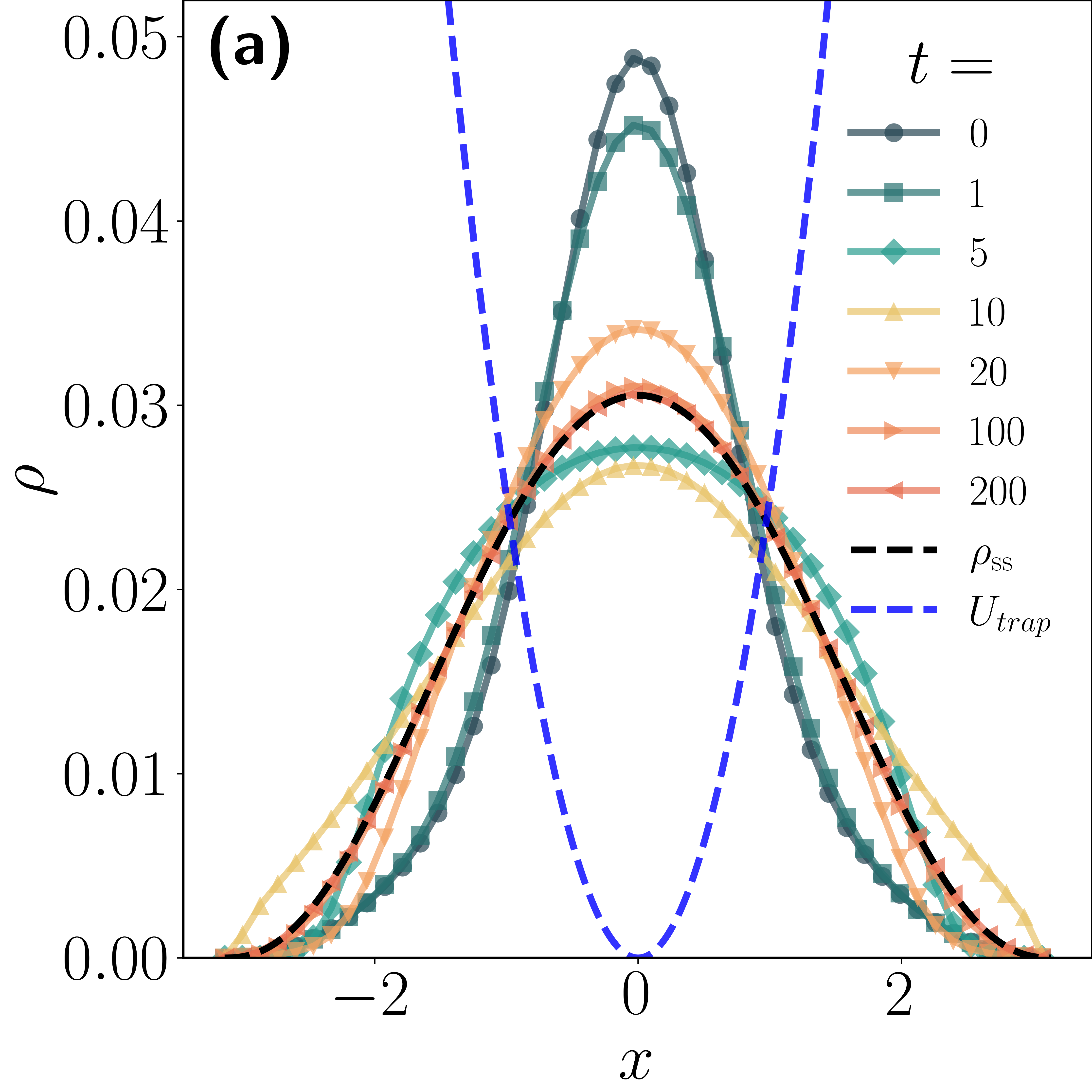}
\includegraphics[width=0.32\linewidth]{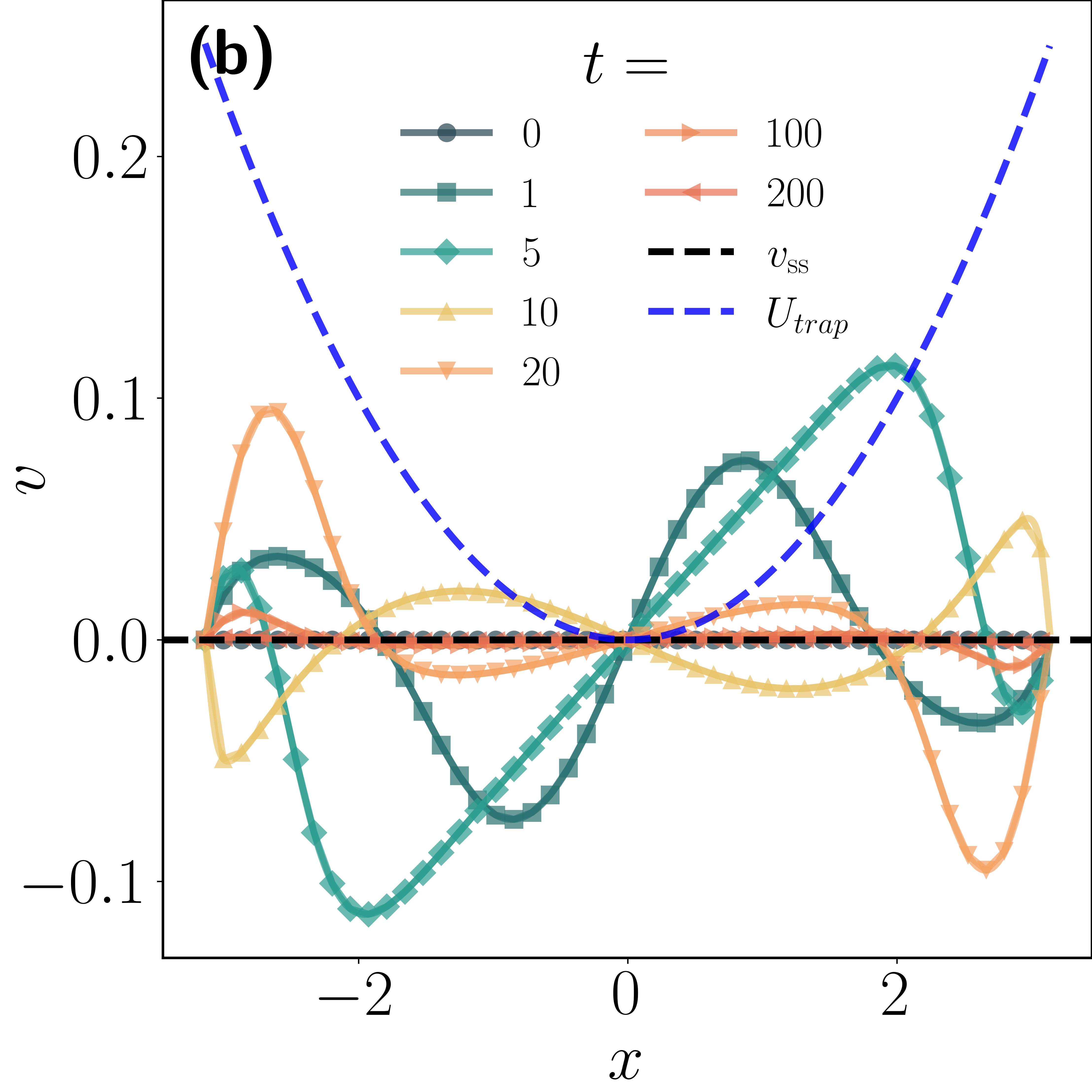}
\includegraphics[width=0.32\linewidth]{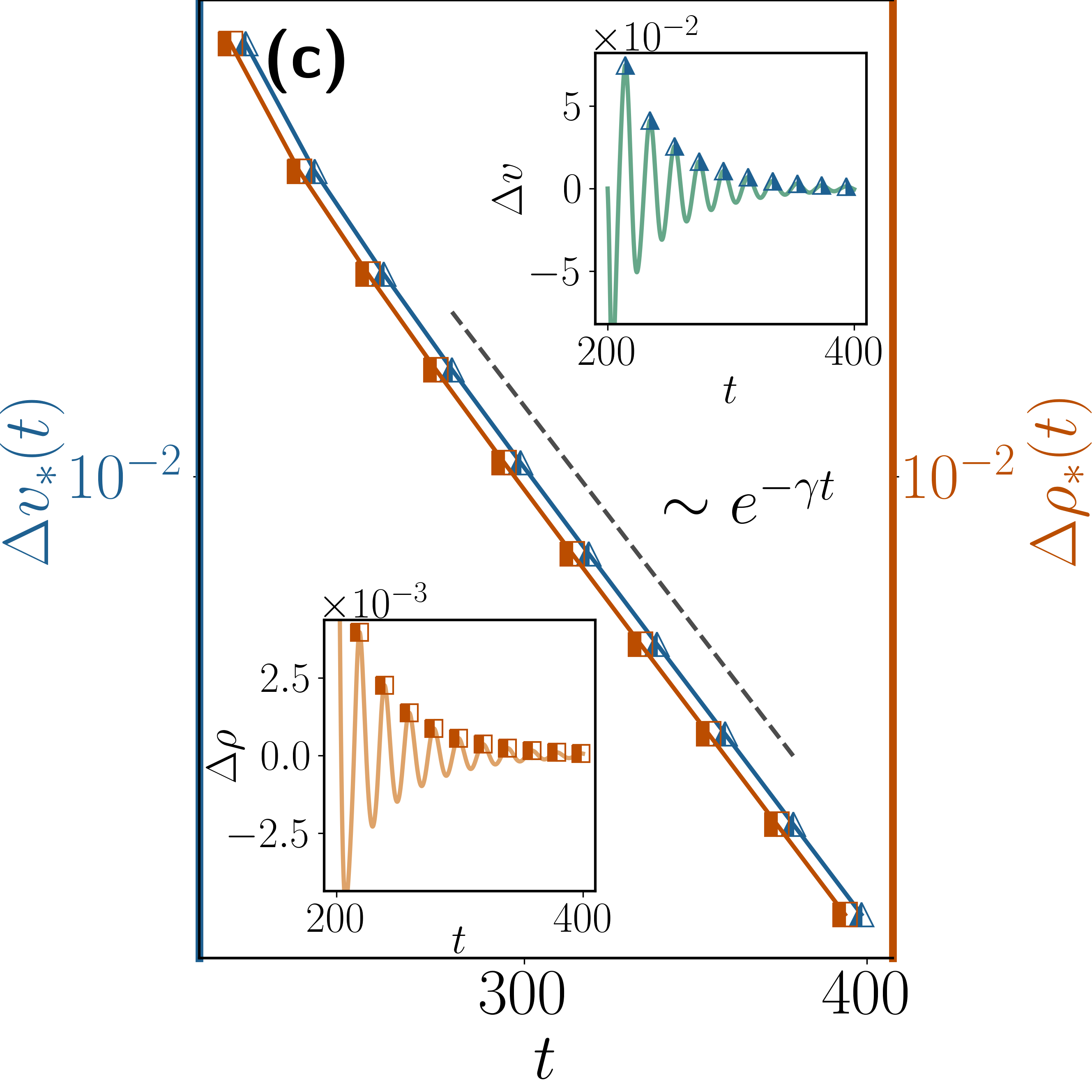}
	\caption{
    Spatial profiles, at different times, of the (a) density and (b) velocity fields of a gas confined in a trap (dashed 
	blue line) with $\omega^2=0.1$ for the fully nonlinear set of equations in  Eq.~\eqref{eq:1Deuler_trap}. (c) Time series of the perturbation in the density $\Delta \rho$ (lower inset) and velocity $\Delta v$ (upper inset) at a fixed chosen location showing a 
	decaying, oscillatory relaxation to the steady state. The successive peaks $\Delta \rho_*$ and $\Delta v_*$ are indicated by red and blue markers. 
	In the main panel we show a semi-log plot of $\Delta \rho_*$ and $\Delta v_*$ suggesting exponentially damped solutions of the evolution equations of the perturbed field. The decay time scale $1/\gamma$ is in excellent agreement with the time-scale obtained from the linearized analysis (Sec.~\ref{sec:trap}) as shown by the dashed line.} \label{fig:trap}
\end{figure*}

\section{Relaxation of a one-dimensional trapped gas}
\label{sec:trap}

\begin{figure*}
\includegraphics[width=0.32\linewidth]{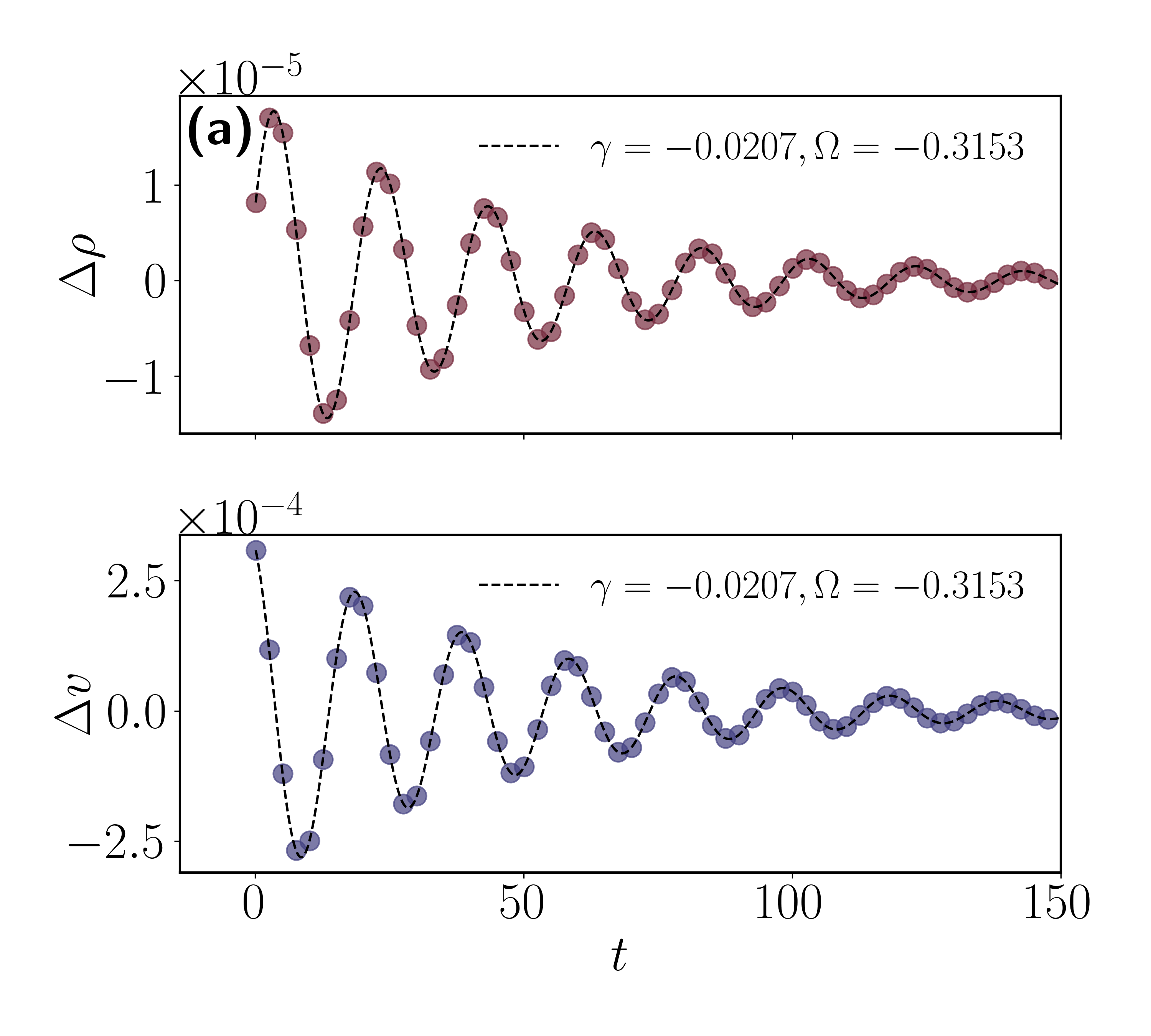}
\includegraphics[width=0.32\linewidth]{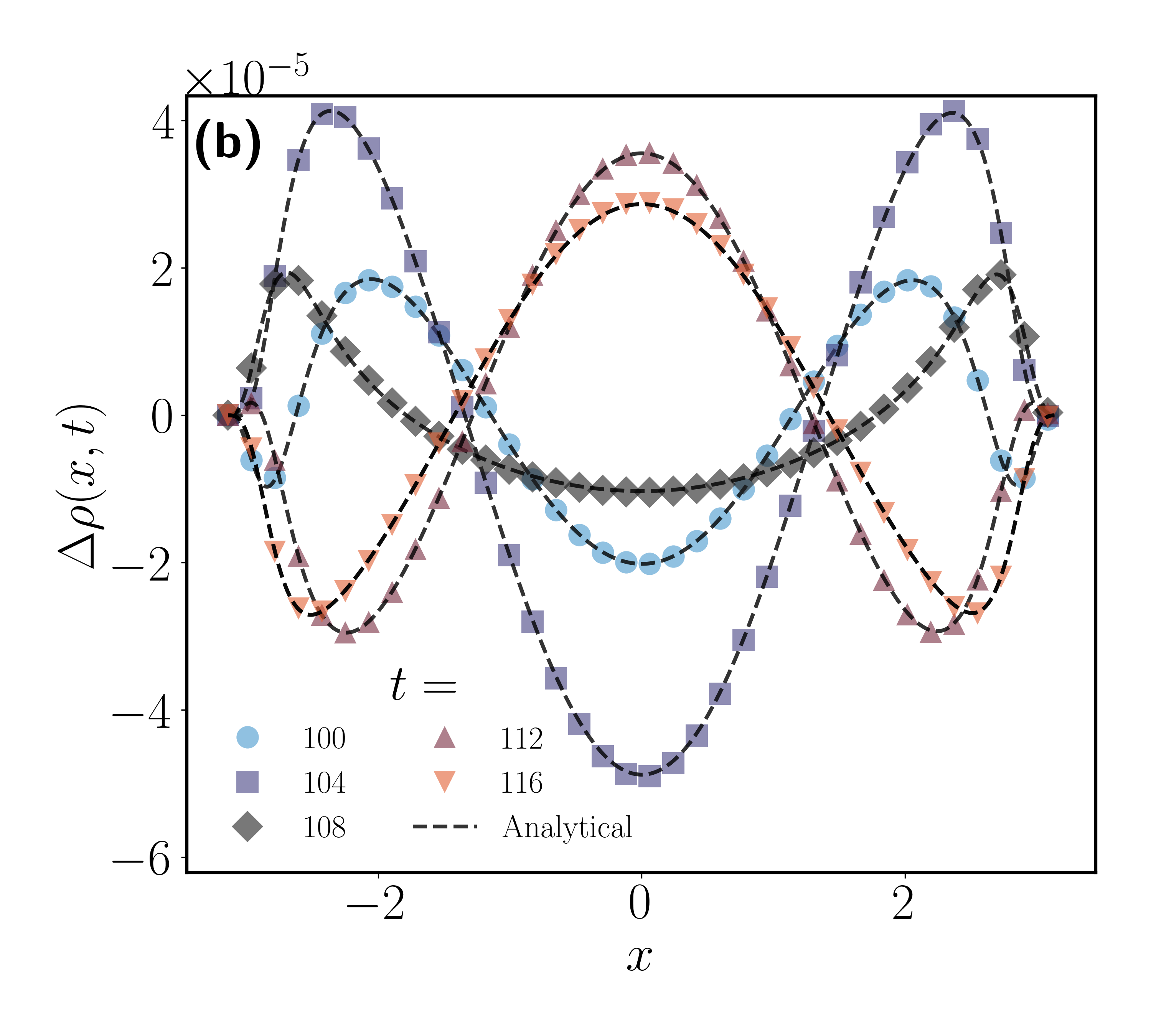}    \includegraphics[width=0.32\linewidth]{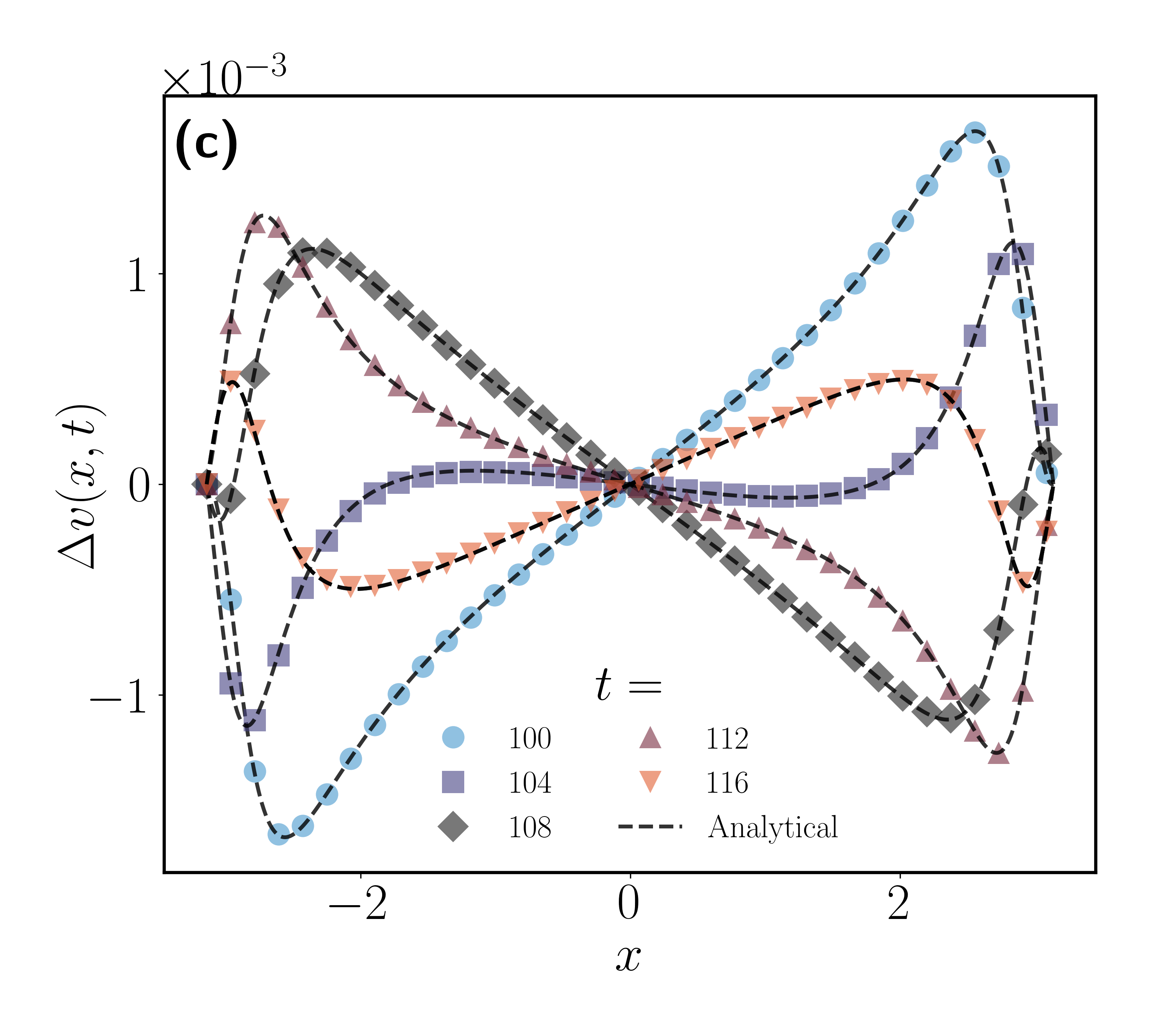}
\caption{Upper and lower half of panel (a) shows the time series of $\Delta \rho$ and $\Delta v$ respectively at a chosen spatial location. The markers shows the data obtained from direct numerical simulation of Eq.~\eqref{eq:pert} and the dashed lines show the fit with a damped sinusoidal function of the form $ C\sin(\Omega\,t+D)e^{-\gamma t}$. The legend shows the extracted fitting parameters namely $\gamma,\Omega$ and this agrees perfectly (up-to four decimal places) with the values obtained by using the boundary conditions given in  Eq.~\eqref{eq:psi_bc}. Panels (b) and (c) shows the perturbed density and velocity profiles for different time snapshots computed by solving the linearized Eq.~\eqref{eq:pert} represented by markers. The dashed line shows the analytical expressions given in Eqs.~\eqref{eq:del_rho} and \eqref{eq:del_v} which perfectly agrees with the numerical simulations of linearised equations.}
\label{fig:lineqcomp}
\end{figure*}

\begin{figure*}
\includegraphics[width=1\linewidth]{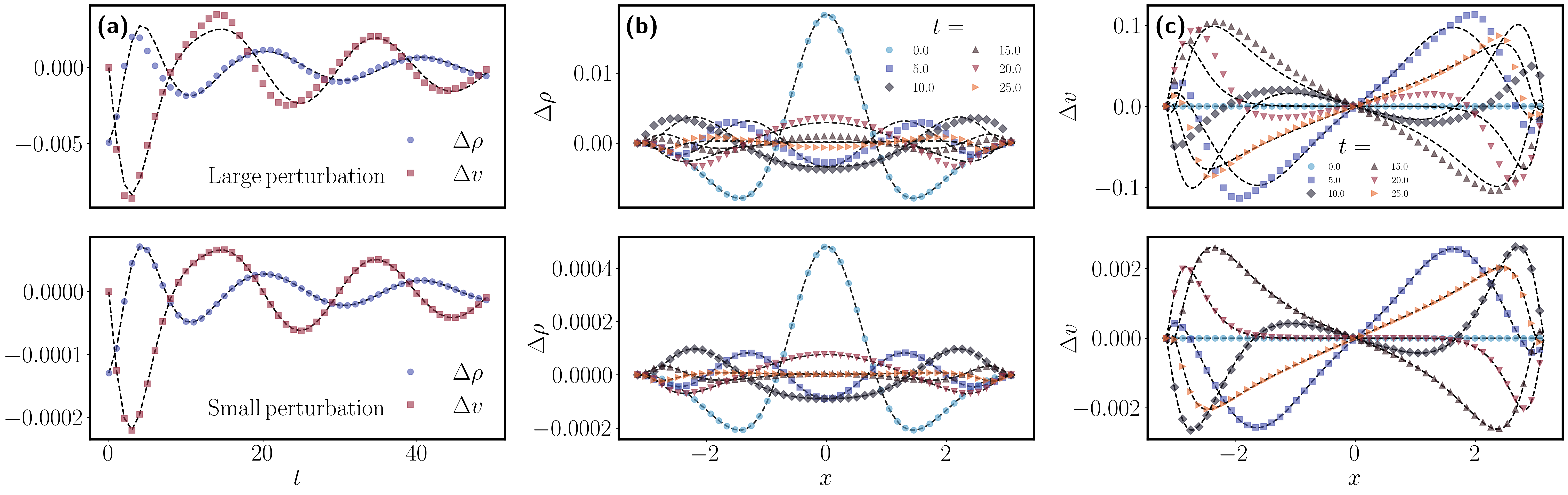}

\caption{(a) Time series of the density $\Delta \rho$ and velocity perturbations $\Delta v$ at a fixed spatial location for large (upper panel) and small (lower panel) initial perturbations. Markers represent the nonlinear evolution [Eq.~\eqref{eq:1Deuler_trap}], while the black dashed line indicates the corresponding linear evolution [Eq.~\eqref{eq:pert}]. Since $\Delta v$ differs significantly in magnitude from $\Delta \rho$, it has been scaled by a constant factor for visualization on the same plot. Panels 
(b) and (c) show the comparison of the full nonlinear and linearized dynamics for the time evolution of $\Delta \rho$ and $\Delta v$ for large (upper panel) and small (lower panel) perturbations. Markers denote the nonlinear evolution, while dashed lines represent the solutions of the linearized equations. Upper panels of (b) and (c) clearly demonstrates that for large perturbations, the linearized and nonlinear evolutions do not agree, whereas the lower panel of (b) and (c) show that for smaller perturbations the linear approximation is excellent agrees with the nonlinear evolution. }
\label{fig:pert_small_large}
\end{figure*}

So far, we have discussed cases when there is no external confinement. In this section, we discuss the case when we have a confining potential $U_{\rm trap} \neq 0$. 
Starting from some nonequilibrium initial condition we study how the gas evolves and eventually reaches a steady state.
The hydrodynamic equations are given by
\begin{subequations} 
		\label{eq:1Deuler_trap}
	\begin{align}
\frac{\partial \rho}{\partial t} + \frac{\partial \rho v}{\partial x} &= 0,
		\label{eq:1Deulerb_trap}
		\\
\frac{\partial v}{\partial t} + \frac{\partial }{\partial x}\left (\frac{v^{2}}{2}+\rho^\delta +\frac{\omega^2}{2} x^2 \right) &= \nu \frac{\partial^2 v}{\partial x^2}, 
		\label{eq:1Deulera_trap}		
	\end{align}
\end{subequations} 
where we are using dimensionless variables as in  Eq.~\eqref{eq:DL} and  $\omega$ and $\nu$ are also made dimensionless via
\begin{align}
\frac{\omega\, \sigma}{(D \rho_0^\delta)^{1/2}} &\to \omega,~~
\frac{\nu}{(D \rho_0^\delta)^{1/2}\sigma}  \to \nu. \label{eq:2dtrap}
\end{align}
Note that a more realistic form for the dissipative term in Eq.~\eqref{eq:2dtrap} would be $\partial_x (\rho \partial_x v)/\rho$. For simplicity, we replace it with the form $\nu \partial^2_x v$ which is expected to be a good approximation for shallow traps and avoids numerical issues related to divergences associated with the $1/\rho$ factor.  

It is easy to see that the steady state solutions of Eq.~\eqref{eq:1Deuler_trap} takes the form
\begin{equation}
	\rho_{\rm ss} = \left(\frac{\omega^2 a^2}{2}\right)^{1/\delta} \left(1 - \frac{x^2}{a^2}\right)^{1/\delta},~~~v_{\rm ss}=0 
	\label{eq:rho_s}
\end{equation}
where the constant $a$ is fixed by the constraint 
\begin{equation}
\int_{-a}^a dx\,\rho_{\rm ss}(x)=M\, .
\end{equation}
In Fig.~\ref{fig:trap}, we show results from direct numerical simulations of Eq.~\eqref{eq:1Deuler_trap} for $\delta=2/5$. Since we are using a pseudospectral method with Fourier basis which is suited for periodic boundaries we smoothen the quadratic potential near the boundary and make it periodic to avoid numerical errors (see Appendix.~\ref{sec:dns}). In other words, without the smoothening, $\frac{\partial U_{trap}}{\partial x}$ is discontinuous at the boundary making the pseudospectral method ill-suited. We show in Fig.~\ref{fig:trap}(a) density and (b) velocity profiles at several different times. Clearly at late time both the density and velocity 
fields converge to analytical expressions given in Eq.~\eqref{eq:rho_s}. Figure~\ref{fig:trap}(c) displays an exponential decay in $\rho$ and $v$ at a fixed spatial location. We analytically derive the corresponding decay timescale later in this section.

As can be noticed in  Fig.~\ref{fig:trap} (a) and (b), both the density and velocity profiles show damped oscillatory relaxation to the steady state. We now provide an in-depth understanding of this transient behaviour and, in particular, compute the associated damping rate and frequency.  A full solution of the time-dependent hydrodynamic equations is challenging. However, for initial conditions that are close to the steady state one can consider a linearized dynamics. As we shall now show, such equations are amenable to analytic treatment and in fact also provide insights into the late time dynamics of even far-from equilibrium initial conditions.   

To proceed further, we make the following substitution into Eq.~\eqref{eq:1Deuler_trap}
\begin{equation}
\rho = \rho_{\rm ss}+\Delta\rho\,, v=\Delta v\, , 
\end{equation}
where recall that $\rho_{\rm ss}$ is given in Eq.~\eqref{eq:rho_s}.  Retaining terms linear in the perturbation we obtain
\begin{subequations} 
	\begin{align}
\frac{\partial \Delta \rho}{\partial t} + \frac{\partial}{\partial x} \rho_{\rm ss} \Delta v &= 0,
		\label{eq:perta} \\
                \frac{\partial \Delta v}{\partial t} + \delta \frac{\partial}{\partial x}  \rho_{\rm ss}^{\delta -1} \Delta \rho &= \nu \frac{\partial^2 \Delta v}{\partial x^2}.
		\label{eq:pertb}
		\end{align}
		\label{eq:pert}
\end{subequations}
Note that Eq.~\eqref{eq:pert} is linear in $\Delta \rho$ and $\Delta v$ but with space dependent coefficients making them still challenging to solve. Nevertheless we can in fact obtain the spectrum completely. 
Let us look for solutions of Eq.~\eqref{eq:pert} of the form
\begin{equation}
(\Delta \rho(x,t),\Delta v(x,t))=(\phi(x),\psi(x)) e^{-\lambda t}.
\end{equation} 
Plugging this into Eq.~\eqref{eq:pert} gives us the following  equations for $\psi$ and $\phi$:
\begin{align}
 & \nu \frac{d^2 \psi}{d x^2}-\frac{\delta}{\lambda} \frac{d}{d x}  \rho_{\rm ss}^{\delta -1} \frac{d  (\rho_{\rm ss} \psi)}{d x} +\lambda {\psi}  =0,  \label{eq:psi} \\
& \phi(x)=\frac{1}{\lambda}\frac{d(\rho_{\rm ss} \psi)}{dx}.
\label{eq:ansatz}
\end{align}
Plugging in the explicit form of $\rho_{\rm ss}$  in Eq.~\eqref{eq:rho_s} into Eq~\eqref{eq:psi} leads to the form:
\begin{equation}
    (c_1+c_2 x^2) \frac{d^2\psi}{dx^2} + c_3 x \frac{d\psi}{dx} + c_4 \psi = 0 \, ,
    \label{eq:psi_legendre}
\end{equation}
where, 
\begin{eqnarray}
    c_1 &=& -\nu + \frac{\omega^2 a^2 \delta}{\lambda}, \,\,   c_2 = -\frac{\omega^2 \delta }{2\lambda },\\
    c_3 &=&-\frac{\omega^2}{\lambda }\left( \delta + 1\right), \,\,  c_4 = - \left( \lambda + \frac{\omega^2}{\lambda}\right).
\end{eqnarray}
These equations have to be solved with the boundary conditions 
\begin{align}
\psi(a)=\psi(-a)=0. 
\label{eq:psi_bc}
\end{align}
This will give us to a set of  eigenfunctions $\{\psi_n,\phi_n\}$ and corresponding allowed eignevalues $\{\lambda_n\}$ for our linear equations. From the structure of the eigenvalue equations [Eq.~\eqref{eq:ansatz}], it is clear that the eigenvalues and eigenfunctions come in complex conjugate pairs. We now note that Eq.~\eqref{eq:psi_legendre}
is in the form of the Legendre equation whose general solution can be expressed in terms of the associated Legendre functions of the first and second kinds:
\begin{equation}
    \psi(x) = R(x)\big[A_1 P(a_1,a_2,a_3x) + A_2 Q(a_1,a_2,a_3x)\big],
\end{equation}
where 
\begin{eqnarray}
    R(x)&=& (c_1 + c_2 x^2)^{-\frac{2c_2+c_3}{4c_2}} \\
    a_1 &=&  \frac{\sqrt{c_2^2-2 c_2 c_3-4 c_2 c_4+c_3^2}-c_2}{2 c_2} \\
    a_2 &=& -\frac{2 c_2 + c_3}{2 c_2}, \quad a_3 = i \sqrt{\frac{c_2}{c_1}}, 
\end{eqnarray}
and $A_1,A_2$ are arbitrary constants. Without loss of generality we can set $A_1=1$ since this can be fixed by normalization. We then have two complex unknown parameters, namely $A_2$ and $\lambda$, These are determined by imposing the two  boundary conditions $\psi_n(\pm a)=0$. We solve for these unknown parameters numerically using Mathematica.  

The general solution of Eq.~\eqref{eq:pert} is given by
\begin{align}
    (\Delta \rho (x,t),\Delta v(x,t))= \sum_{n=0}^\infty p_n [\phi_n(x), \psi_n(x)] e^{-\lambda_n t}, 
    \label{eq:rho_v_exp}
\end{align}
where the constants $p_n$ are determined from the initial conditions. At large times,  the eigenvalue with the smallest value for the real part will dominate. There are two such eigenvalues  which we denote  as 
\begin{equation}
\lambda_m=\gamma \pm i \Omega,
\label{eq:lam}
\end{equation}
where $\gamma>0$ and $\Omega$ are real. These provide us with the time scale for decay and the frequency of the oscillations, namely $\gamma^{-1}$ and $\Omega$ respectively. The expected forms for the density and velocity fields at late times can also be determined. Writing the dominant eigenfunctions [Eq.~\ref{eq:rho_v_exp}] in the form 
\begin{equation}
\phi_m(x)=F_1(x)e^{\pm i\alpha_1(x)} ,\quad  \psi_m(x)=F_2(x) e^{\pm i\alpha_2(x)}\,,
\end{equation}
we have the following late time forms:
\begin{align}
\label{eq:del_rho}
    \Delta \rho(x,t)&= C_1 F_1(x) \sin(\alpha_1(x)-\beta_1- \Omega \, t) e^{-\gamma t},\\ \Delta v(x,t)&= C_2 F_2(x) \sin(\alpha_2(x)-\beta_2-\Omega \, t) e^{-\gamma t}, 
\label{eq:del_v}
\end{align}
where $C_1,C_2,\beta_1,\beta_2$ are  constants that are fixed by the initial conditions. In the numerics for the nonlinear case these constants  can be fixed by using the value of $\Delta \rho$ and $\Delta v$ at some late time snapshot because in the early time there might be contribution from other eigen-functions which may lead to an incorrect estimation of the constants. In Fig.~\eqref{fig:lineqcomp}, we compare the analytical predictions [Eqs.~\eqref{eq:del_rho}, \eqref{eq:del_v}] with direct numerical solutions of the linearized equations [Eq.~\eqref{eq:pert}] and find excellent agreement. To further validate the calculations with the linearized equations, in Fig.~\ref{fig:pert_small_large} we explicitly compare numerical simulations of the full nonlinear system [Eq.~\eqref{eq:1Deuler_trap}] with those of the perturbation equations [Eq.~\eqref{eq:pert}]. For small perturbations, the linearized equations accurately approximate the full nonlinear dynamics. However, as the perturbation strength increases, deviations become apparent, indicating the breakdown of the linear approximation.
 \section{Two-dimensional case}
\label{sec:2d}
\begin{figure*}
    \includegraphics[width=0.97\linewidth]{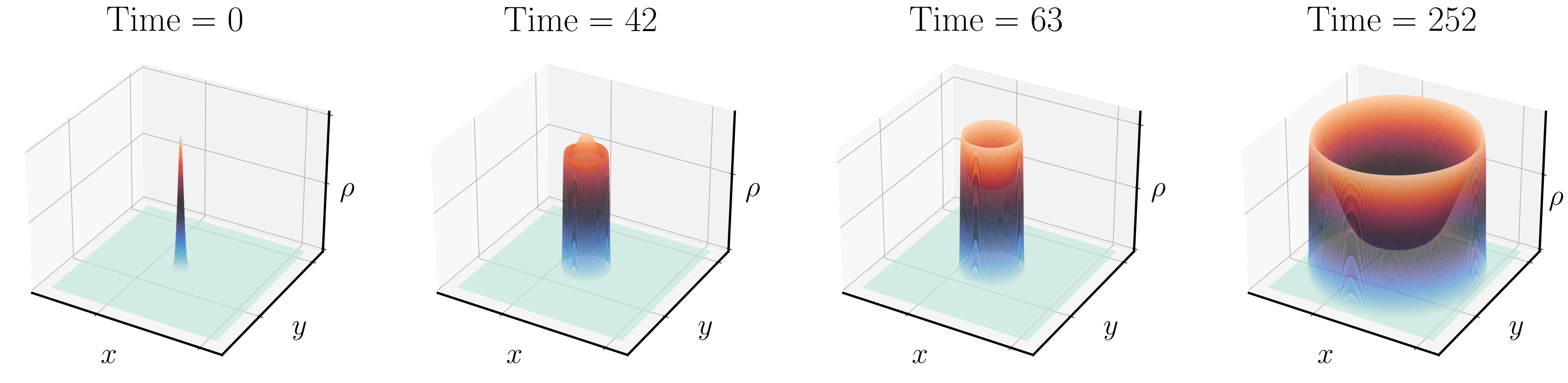}
    \caption{The time evolution of the two-dimensional density for $\delta = \frac{2}{5}$  starting from a localized density profile and zero momentum, we clearly observe the emergence of a shock-like structure and a scaling form. Animation of the time evolution is available in Ref~\onlinecite{git_Ritwik}.
}
\label{fig:2d_full}
\end{figure*}

\begin{figure*}
    \includegraphics[width=0.98\linewidth]{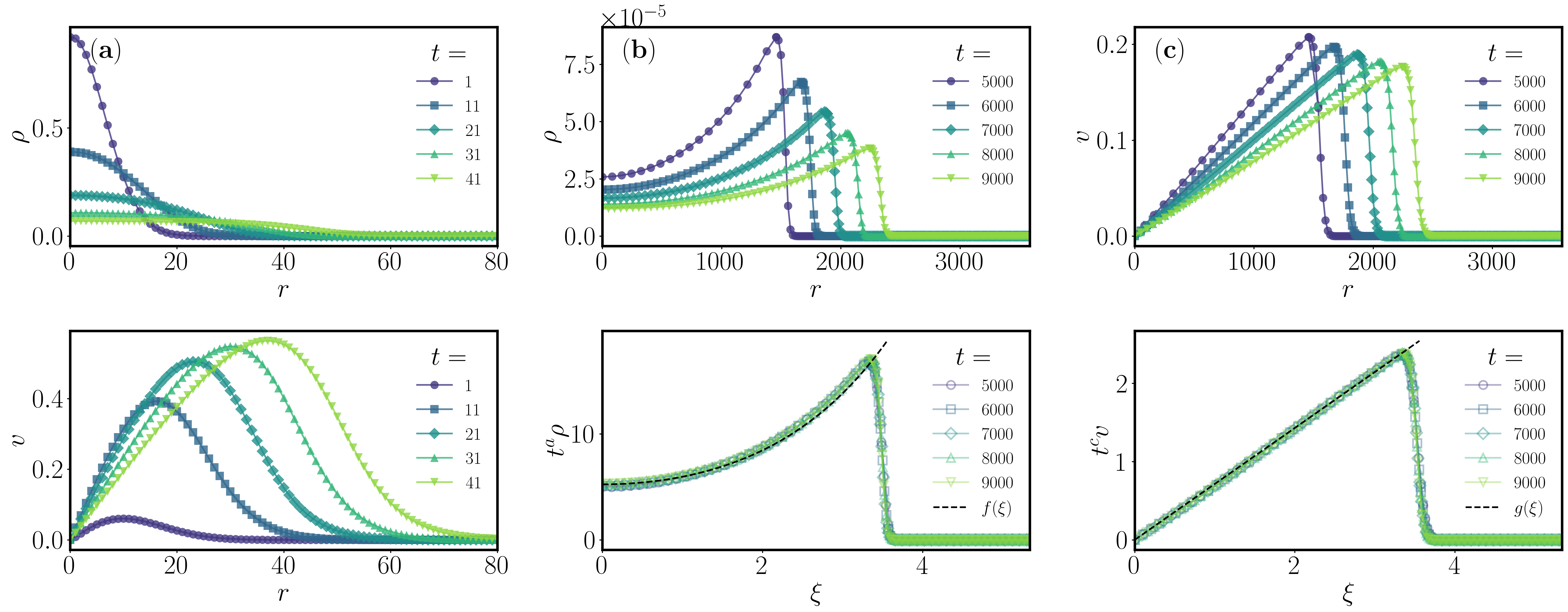}
    \caption{(a) The early time evolution of the radial density and the velocity before developing the scaling form as described in Eq.~\eqref{eq:app:2d} for $d=2$ and $\delta=2/5$.  Upper panel of (b) and (c) show the radial density and velocity profiles for late times after development of the scaling form whereas the lower half of panel (b) and (c) show the rescaled density and velocity profiles and the analytically computed scaling functions $f$ and $g$ described in Eq.~\eqref{eq:2d_fg}. Note that the shock is smoothened; hence the location of the shock front used in Eq.~\eqref{eq:2d_fg} is the mean of this smoothened shock front which leads to a difference of 1\% from the analytical obtained shock position in Eq.~\eqref{eq:front}.}
    \label{fig:2d}
\end{figure*}

In this section, we discuss the evolution, in two dimensions, of an initially localized mass of a quantum gas  with zero total momentum. We only consider expansion into a vacuum. The initial mass distribution is assumed to have radial symmetry and so we look for solutions of the hydrodynamic equations which have this symmetry. In particular, we need to only consider the density field $\rho$, and the radial component of the 
velocity vector field $v$. Clearly, the space-time dependence of the fields will have the form $\rho=\rho(r,t), v=v(r,t)$ and the corresponding Euler equations in radial coordinates are
\begin{subequations}
	\begin{align}
	\frac{\partial \rho}{\partial t}  + \frac{\partial \rho v}{\partial r}  + \frac{d-1}{r}\rho v_r = 0  
		\label{eq:app:2d_rho} \\
            \frac{\partial v}{\partial t}  + \frac{\partial }{\partial r} \left( \frac{v^2}{2}+  \rho^{\delta}\right)   = 0.
		\label{eq:app:2d_v} 
	\end{align}
	\label{eq:app:2d}
\end{subequations}
As in Sec.~\ref{sec:vac} we expect that asymptotically the dynamics develops a scaling form as
\begin{equation}
	 \rho (r,t) = {t^{-2b}} \, f(\xi),\,\, 
	v(r,t) = t^{-c}  g(\xi),\,  
    {\rm with}~~~\xi =\frac{r}{t^b},
	\label{eq:scaled_2d}
\end{equation}
where the factor, $2b$, for the density scaling appears since $\int_0^\infty dr r \rho(r,t)$ has to be conserved. 
Plugging Eq.~\eqref{eq:scaled_2d} into Eq.~\eqref{eq:app:2d} and requiring that any explicit time dependence goes away, we get
\begin{equation}
	\label{2d_scaling} 
	  b = \frac{1}{1+\delta} \quad , \quad
	c = \delta\, b
\end{equation}
and the corresponding ordinary differential equations for the scaling functions
\begin{subequations} 
\begin{align}
	 (g-b\xi)\frac{df}{d\xi} + f \frac{dg}{d\xi} + \frac{fg}{\xi} - 2 b f &= 0; 
\label{eq:app:2Dodea}\\
    \frac{df^\delta}{d\xi} + (g - b\xi) \frac{dg}{d\xi}  -cg &= 0.
\label{eq:app:2Dodeb}
\end{align}
\label{eq:app:2Dode}
\end{subequations} 
We now discuss the Rankine-Hugoniot conditions which turn out to be identical to the 1D case given by Eqs.~\eqref{eq:rh} or \eqref{eq:main:rh_cond}. To derive the first Rankine–Hugoniot condition, we integrate the first conserved quantity, $\int r \rho \, {\rm d}r$, across the shock at position $R$, i.e., over the interval $[R - \Delta, R + \Delta]$ and then take the limit as $\Delta \to 0$. This yields
\begin{eqnarray}
        \frac{d}{dt} \int_{R-\Delta}^{R+\Delta}r\rho \, {\rm d}r 
    &=& 0
\end{eqnarray}
which implies
\begin{eqnarray}
       (\rho_{-} - \rho_{+})RU - ( \rho_{-}v_{-} - \rho_{+}v_{+}) R &=& 0\, ,
\end{eqnarray}
where $\rho_{-}$ and $v_{-}$ denote the radial density and velocity before the shock, and $\rho_{+}$ and $v_{+}$ denote the density and velocity after the shock. Noting that $\rho_{+} = v_{+} = 0$ gives the first Rankine–Hugoniot condition
\begin{equation}
    v_{-} = U.
    \label{eq:2d_RH1}
\end{equation}
To derive the second Rankine–Hugoniot condition, we use the second conserved quantity, $\int rv \, \mathrm{d}r$, and integrate it around the shock
\begin{eqnarray}
        \frac{d}{dt} \int_{R-\Delta}^{R+\Delta}rv \, {\rm d}r 
    &=&  0 .
    \label{eq:RH-inter}
\end{eqnarray}
which implies
\begin{eqnarray}
        (r_{-}v_{-}-r_{+}v_{+})U + \int_{R-\Delta}^{R+\Delta} r\partial_t v \, {\rm d}r
    &=& 0.
    \label{eq:RH-inter}
\end{eqnarray}
We now simplify the second term
\begin{eqnarray}
\label{eq:simpint}
      \int_{R-\Delta}^{R+\Delta} r\partial_tv \, {\rm d}x 
      &&= - \int_{R-\Delta}^{R+\Delta} r\partial_r\left(\frac{1}{2}v^2 + \rho^\delta\right)\, {\rm d}r \\\nonumber
      && = \int_{R-\Delta}^{R+\Delta} \bigg[ \left(\frac{1}{2}v^2 + \rho^\delta\right) \\ \nonumber &&~~~~~~~~~~~~~-  \partial_r\left(\frac{1}{2}rv^2 + r\rho^\delta\right)  \bigg]{\rm d}r.\\ \nonumber
\end{eqnarray}
Plugging Eq.~\eqref{eq:simpint} back in Eq.~\eqref{eq:RH-inter} and taking the limit $\Delta \to 0$ gives us the second Rankine-Hugoniot condition
\begin{equation}
v_{-}U = \frac{1}{2}v_{-}^2 + \rho_{-}^{\delta}.
\label{eq:2d_RH2}
\end{equation}
By using Eq.~\eqref{eq:2d_RH1} in Eq.~\eqref{eq:2d_RH2}, we get
\begin{equation}
    \rho_{-}^\delta = \frac{1}{2}U^2.
\end{equation}
In terms of scaled variables given in Eq.~\eqref{eq:scaled_2d}, we get 
\begin{equation}
    g_{-} = b\,\xi_f ,\quad f_{-} = \left( \frac{1}{2}b^2\xi_f^2\right)^{1/\delta},
    \label{eq:2d_xi_f}
\end{equation}
which takes into account that the shock position is given by $R=\xi_f t^b$ and shock velocity $U=\dot{R}$. 

We now want to find the solutions to Eq.~\eqref{2d_scaling}. We take an \textit{Ans\"atz} $g(\xi) = k \,\xi$ and substitute it in Eq.~\eqref{eq:app:2Dodea} yielding
\begin{equation}
\label{eq:sub1}
    (-b\xi + k\xi) \frac{df}{d\xi} + kf - af +kf = 0.
\end{equation}
To ensure a non-zero solution for Eq.~\eqref{eq:sub1}, we need $k=b$. So, $g(\xi)=b\,\xi$. We verify via direct numerical simulations that this is indeed the case. Now we substitute $g(\xi)=b\,\xi$ in Eq.~\eqref{eq:app:2Dodeb} and get
\begin{eqnarray}
    \delta f^{\delta-1} \frac{df}{d\xi} - c \,b\, \xi &=& 0,
\end{eqnarray}
which simplifies to
\begin{eqnarray}
    \frac{d f^{\delta}}{d\xi} = c\,b\,\xi.
    \label{eq:fsimp}
\end{eqnarray}
Integrating Eq.~\eqref{eq:fsimp} from $\xi_f$ to $\xi$ we get,
\begin{equation}
f^{\delta}(\xi) = \left(f^{\delta}(\xi_f) - \frac{c\,b}{2}(\xi_f^2-\xi^2)\right)^{1/\delta}.
\end{equation}
Finally, the analytical solution for $f$ and $g$ can be summarized as
\begin{equation}
    f^{\delta}(\xi) = \left(f^{\delta}(\xi_f) - \frac{c\,b}{2}(\xi_f^2-\xi^2)\right)^{1/\delta},\quad g = b\,\xi 
    \label{eq:2d_fg}
\end{equation}
where we recall that $b$ is given in Eq.~\eqref{2d_scaling} and $f(\xi_f)$ can be obtained from the 
Rankine-Hugoniot condition [Eq.~\eqref{eq:2d_xi_f}]. What remains is to find $\xi_f$ from mass conservation
\begin{align}
\int_0^{\xi_f} d\xi \,\xi f(\xi)=M,
\end{align}
which yields (for $\delta<1$) 
\begin{equation}
    \xi_f = \left(\frac{ \left(1-(1-\delta )^{\frac{1}{\delta }+1}\right) (\delta +1)^{-\frac{\delta +2}{\delta }}}{2^{\frac{\delta +1}{\delta }} M}\right)^{-\frac{\delta }{2 \delta +2}}\, .
    \label{eq:front}
\end{equation}

Next we test our analytical prediction via direct numerical simulations. First we solve the full two-dimensional equations i.e. Eqs~\eqref{eq:cont} and~\eqref{eq:euler} for $\delta=2/5$ and $U_{trap}=0$ starting from a localised density profile and zero velocity. Figure.~\ref{fig:2d_full} shows the density evolution for early and late times demonstrating blast-like phenomenon and the emergence of a scaling form. In Fig.~\ref{fig:2d}, we show the numerical results obtained from solving Eq.~\eqref{eq:app:2d} numerically for $d=2$ and $\delta=2/5$. (a) Figure~\ref{fig:2d}(a) shows the early time behavior and upper half of panels (b) and (c) show the radial density and velocity for relatively late times. In the lower panels pf Fig.~\ref{fig:2d} (b) and (c) we show excellent agreement between rescaled numerical profiles and analytical scaling functions given in Eq.~\eqref{eq:2d_fg}.

\section{Conclusions and Outlook}
\label{sec:conc}

To summarize, we have investigated quantum dynamics in  highly out-of-equilibrium situations, both in one and two dimension,  via hydrodynamics and using both  analytical calculations and direct numerical simulations. We demonstrated that at long times, the hydrodynamic fields evolve to self-similar forms, that are independent of the details of the localized initial conditions.  We derived analytically the scaling exponents and scaling functions and thus characterized  completely the features of this highly nonequilibrium process. In particular, for a gas with the equation of state $P\sim\rho^{\delta+1}$ and evolving in a vaccuum, the shock front  grows with time as $R(t) \sim t^b$, where $b=2/(2+\delta)$ and $1/(1+\delta)$ in one and two dimensions respectively. 
We also investigated the relaxation dynamics of a confined quantum gas. At  late times we see a damped oscillatory relaxation, where the damping coefficient and frequency can be computed analytically from a linearized theory. We validated our analytical findings by extensive numerical simulations and  found excellent agreement between the two. 

Our methodology is highly adaptive and expected to be of relevance in understanding far from equilibrium dynamics in a variety of interacting quantum systems. 
It will be interesting to explore quantum gases in higher spatial dimensions, relaxing the assumptions of symmetry~\cite{ketterle2008making,Hou_anisotroy,Salasnich_anisotroy}, which would make the effective one dimensional description ill-suited. It would also be important and interesting to extend our techniques to open quantum systems such as dissipative quantum fluids~\cite{amo_science,Carusotto_RMP}.
Beyond cold quantum gases, our collective formalism and subsequent analysis of self-similar solutions is also relevant in several other contexts such as finite-ranged Riesz gas~\cite{kumar-frg} and hydrodynamics of a plasma of strongly
interacting quarks and gluons produced in relativistic heavy ion collisions~\cite{SHURYAK200948, Schäfer_2009,dt-qcd,Adams_2012}.
 As noted earlier, our protocol is well-suited to TOF measurements, where atoms are cooled, released, and imaged via absorption~\cite{abs_img_1,abs_img_2,abs_img_3}. Testing our predictions is thus  highly feasible and the observation of self-similar scaling solutions seems like a fascinating possibility. 

\section*{Acknowledgements}
	We thank P. L. Krapivsky for very helpful discussions. AD and MK thank the VAJRA faculty scheme (No.
VJR/2019/000079) from the Science and Engineering Research Board (SERB), Department of Science and Technology, Government of India. AD acknowledges the J.C. Bose
Fellowship (JCB/2022/000014) of the Science and Engineering Research Board of the Department of Science and Technology, Government of India. AD and MK thank the hospitality of Yukawa Institute for Theoretical Physics, Kyoto University during the workshop on ``Hydrodynamics of low-dimensional interacting systems: Advances, challenges, and future direction". 
    The authors 
	acknowledge the support of the DAE, Government of India,
	under projects nos. 12-R\&D-TFR-5.10-1100 and RTI4001.
\appendix

\section{Direct Numerical Simulations} 
\label{sec:dns}
We perform direct numerical simulations using a fully-dealiased pseudospectral method with periodic boundary conditions in both one and two dimensions. Spatial discretization is carried out using a Fourier basis, and time integration is performed via either the classical fourth-order Runge--Kutta (RK4) method or the integrating factor RK4 (IFRK4) scheme, depending on the stiffness of the equations~\cite{cox_matthews}.

To simulate the inviscid equations, we introduce a small artificial viscosity in both the density and velocity equations to regularize sharp gradients and suppress numerical instabilities. More precisely, in the simulations we replace the Euler equations in Eqs.~\eqref{eq:cont}-\eqref{eq:euler} and work with the following dissipative equations 
\begin{subequations}
\begin{align}
\label{eq:cont_num}	
	&\frac{\partial \rho}{\partial t} + \nabla\cdot(\rho {\bf v}) = \eta \frac{\partial^2 \rho}{\partial x^2}, \\
    &\frac{\partial {\bf v}}{\partial t} + {\bf v}\cdot\nabla {\bf v} + \nabla w +\nabla U_{\rm trap}=   \nu \frac{\partial^2 v}{\partial x^2},
\label{eq:euler_num}	
\end{align}
\end{subequations}
For the range of values that we consider in this paper, i.e, $0<\delta<2$ in 1D and $0<\delta <1$ in 2D, one can easily check that, for the case of free evolution, the dissipation terms become irrelevant in the long time scaling regime. The viscosity is chosen to ensure numerical stability while preserving the key features of the inviscid dynamics. Convergence is verified by simultaneously increasing the resolution and decreasing the viscosity~\cite{Murugan_PRR}; in the asymptotic regime, we confirm that the features of the solution remain unaffected for the case of free evolution. Note however, that for the 1D trapped gas discussed in Sec.~\ref{sec:trap}, the viscosity plays a significant role in the long time relaxation to the steady state.

For one-dimensional simulations, we use $2^{12} \le N \le 2^{16}$ collocation points. In two dimensions, we use $N \times N$ grids with $2^9 \le N \le 2^{11}$. The time step, $ dt$, is chosen adaptively based on the system size, subject to the Courant–Friedrichs–Lewy (CFL) stability condition.

For freely evolving systems (i.e., without background density), we set the system size to  $L = N$  ($dx = 1$), and find convergence in the scaling regime for $ L \sim 2^{14} $. For cases with a finite background density, convergence is slower, and we obtain reliable results for $ L \sim 2^{16} $.

In the presence of a harmonic trapping potential, we solve Eq.~\eqref{eq:1Deuler_trap} without adding artificial viscosity to the continuity equation. While the Fourier pseudospectral method remains applicable, the non-periodic nature of the potential is incompatible with strict periodic boundary conditions. To suppress boundary artifacts, we apply a mild smoothing of the potential near the edges of the domain. All simulations are performed with a fixed computational domain of length \( L = 2\pi \). This issue does not arise in the linearized perturbation equations [Eq~\eqref{eq:pert}], where the quadratic trap does not appear explicitly and its effects are encoded through the steady-state background.

To solve the two-dimensional equations in polar coordinates [Eq.~\eqref{eq:app:2d}], we impose non-periodic boundary conditions appropriate for the radial geometry. Specifically, we set $v = 0$ and $\partial_r \rho = 0$ at $r = 0$, following Ref~\onlinecite{kumar2022blast}. In contrast to the Cartesian case, periodic boundary conditions hence pseudospectral methods are no longer applicable. We used a second-order finite difference scheme to approximate the spatial derivatives, while time integration was carried out using the classical fourth-order Runge–Kutta (RK4) method.

Most of the codes used in the numerical results reported here can be found in Ref~\onlinecite{git_Ritwik}.\\

\bibliography{references}
\end{document}